%% file: main.tex
\documentclass[aps,prb,twocolumn,superscriptaddress,floatfix,citeautoscript,noeprint]{revtex4-2}
\usepackage{amsfonts, amssymb, amsmath, amsthm}
\usepackage{array}
\usepackage{bm}
\usepackage{braket}
\usepackage{dcolumn}
\usepackage{multirow}
\usepackage{makecell}
\usepackage{graphicx}
\usepackage{hyperref}
\hypersetup{
  colorlinks,
  citecolor=blue,
  filecolor=blue,
  linkcolor=blue,
  urlcolor=blue
}
\usepackage[nameinlink,capitalize]{cleveref}
\usepackage[normalem]{ulem}
\usepackage{dsfont}
\usepackage{indentfirst}
\usepackage{listings}
\usepackage{subfigure}
\usepackage{titlesec}
\usepackage{color}
\usepackage{xcolor}
\usepackage{enumitem}
\usepackage[title]{appendix}
\usepackage{hhline}
\usepackage{epigraph}
\usepackage{adjustbox}
\usepackage[shortcuts]{extdash}

\def\Id{\mathds{1}}
\def\id{\mathds{1}}

\def\Z2{Z_{\mathrm{st2}}}

\def\dInt{{\mathrm{d}}}

\def\varO{{\mathcal{O}}}
\def\Tr{{\mathrm{Tr}}}
\def\tr{{\mathrm{tr}}}

\def\nn{\nonumber}

\newcommand{\norm}[1]{\left\lVert #1 \right\rVert}
\newcommand{\floor}[1]{\lfloor #1 \rfloor}

\NewDocumentEnvironment{alignb}{b}{%
  \begin{align*}
  \refstepcounter{equation} #1 \tag{\theequation}
  \end{align*}
}{\ignorespacesafterend}

\newcommand{\wt}{\widetilde}
\renewcommand{\tilde}{\wt}

\begin{document}

\title{How thermal is a filtered state?}

\begin{abstract}

Quantum many-body states with sufficiently low energy variance can serve as approximations to thermal states, and they may be prepared by energy filtering simple pure states. In this work, we examine how narrow the filter width must be to guarantee thermal behavior. To this end, we analyze the problem in the Floquet regime, where filtered states are found to be equivalent to time averages. This equivalence allows us to reproduce the distinct Rényi-$\alpha$ entropy scalings as reported in [\citeauthor{Morettini2024}, 
Physical Review Letters 133, 240401 (2024)]. 
Crucially, we show that under the Floquet eigenstate thermalization hypothesis, the trace distance between Floquet-filtered states and thermal states is bounded by the square root of filter width. We further demonstrate that these results extend naturally to the conventional Hamiltonian setting by mapping Hamiltonian-filtered states to their Floquet counterparts.

\end{abstract}

\author{Yilun Yang}
\affiliation{Max-Planck-Institut f\"ur Quantenoptik, Hans-Kopfermann-Str.\ 1, D-85748 Garching, Germany}
\affiliation{Munich Center for Quantum Science and Technology (MCQST), Schellingstr. 4, D-80799 M\"unchen}
\affiliation{Department of Mathematics, University of California, Berkeley, California 94720, USA}
\author{J. Ignacio Cirac}
\affiliation{Max-Planck-Institut f\"ur Quantenoptik, Hans-Kopfermann-Str.\ 1, D-85748 Garching, Germany}
\affiliation{Munich Center for Quantum Science and Technology (MCQST), Schellingstr. 4, D-80799 M\"unchen}	
\author{Mari Carmen Ba\~nuls}
\affiliation{Max-Planck-Institut f\"ur Quantenoptik, Hans-Kopfermann-Str.\ 1, D-85748 Garching, Germany}
\affiliation{Munich Center for Quantum Science and Technology (MCQST), Schellingstr. 4, D-80799 M\"unchen}	

\maketitle

\input{introduction}
\input{setup}

\input{entanglement}

\input{thermality}
\input{discussion}

\bibliography{bibliography}

\appendix
\input{appendix}

\end{document}

%% file: introduction.tex
\section{Introduction}

One major potential application of quantum simulation~\citep{Feynman1982,Lloyd1996} is to probe properties of quantum many-body systems at thermal equilibrium. {In this manuscript, we in particular focus on local Hamiltonians.} Whereas directly preparing thermal states in a quantum device can be difficult, a recently developed approach approximates thermal values from easily preparable initial states with the use of energy filters~\citep{Yang2020,Lu2020,Cakan2021,Yang2022,Luo2024}. The filters can filter out energy eigenstates outside the desired energy interval to achieve microcanonical ensembles. In particular, they can be applied to simple initial states to control energy fluctuations. Such filtered states are expected to be thermal with sufficiently small filter width $\delta$.
Indeed, in the most extreme limit, where $\delta$ is exponentially small in system size, the filter resolves individual eigenstates and thus, under the assumption of eigenstate thermalization hypothesis (ETH)~\citep{Deutsch1991,Srednicki1994,Rigol2008a,Deutsch2018}, filtered states can definitely approximate thermal states.

However, the question of how small the filter width $\delta$ needs to be to achieve such \emph{thermalization} remains open. Previous works have approached this problem from different angles. One of them is canonical universality~\citep{Goldstein2006,Dymarsky2019}, which conjectures that pure states supported within a narrow energy window of width $\delta$ can approximate thermal ensembles with an error polynomial in $\delta$. 
While it demonstrates conceptual links to the well-established ETH framework, the underlying theoretical foundation still warrants further investigation.

Another approach is to monitor the entanglement entropies of filtered states~\citep{Banuls2020,Rai2023,Morettini2024}. 
The growth of subsystem entropy as the filter width is decreased~\citep{Banuls2020,Rai2023,Morettini2024} can also inform the local approach to thermality, since in the thermal state the  entropy generically scales with the volume of the subsystem~\cite{Gemmer2001,Gogolin2016}. 
The relation between the energy variance and the entanglement entropy of pure states obtained by spectral filtering was explored in \cite{Banuls2020}, establishing an upper bound on the von Neumann entanglement entropy that scales linearly with $1/\delta$, for generic local interacting Hamiltonians and initial product states.
This scaling suggests that, for a system of size $N$, it suffices to choose $\delta = \varO(1/N)$ to reach thermal equilibrium. 
The numerical evidence in~ \cite{Banuls2020} showed the distance between filtered and thermal states decreasing as the filter width $\delta$ shrinks with $N$, although accessing the regime $\delta \propto 1/N$ for large systems is prohibitive for classical simulations due to the entanglement barrier~\citep{Schuch2008}. 

Subsequent works refined this analysis. \citet{Rai2023} derived a more rigorous upper bound on the von Neumann entanglement entropy, demonstrating that at most an additional logarithmic factor $\sqrt{\log(N/\delta)}$ appears in the leading $1/\delta$ term. In a different work, \citet{Morettini2024} computed Rényi-$\alpha$ entanglement entropies for local Hamiltonians and uncovered distinct scaling behaviors depending on $\alpha$. Interestingly, for $\alpha > 1$, $S_{\alpha}(\rho_{A,\delta})$ grows only logarithmically with $1/\delta$.
Since in generic thermal states all $S_{\alpha}$ satisfy a volume law, this scaling weakens the arguments for the thermality of filtered states.

These results provide valuable insights but have not directly resolved the underlying question: \emph{How thermal is a filtered state?} 
In this work, we address this challenge by utilizing stroboscopic Floquet dynamics generated through discrete-time evolution as the fundamental building block, rather than continuous-time Hamiltonian evolution. This provides a more general framework that naturally recovers the Hamiltonian limit when the driving frequency is sufficiently large.
Under the generic assumption that return probabilities of product states decay exponentially fast with system size, we show that Floquet filtered states are equivalent to averages of states evolved with Floquet dynamics.
Based on this fact, we are able to reconstruct the reduced density matrices (RDM) of Floquet filtered states from their eigenvalue distributions. 
This approach enables us to compute the R\'enyi entanglement entropies of Floquet filtered states, which we find to follow the same scaling laws established in \cite{Morettini2024}.
Furthermore, as the main result, we show that the trace distance between filtered and thermal states is upper bounded by $\varO(\sqrt{\delta})$, under the assumption of Floquet ETH~\cite{DAlessio2014}. Remarkably, this result can be extended to Hamiltonian filtered states, i.e., states filtered with a Hamiltonian filter, by mapping them to effective Floquet filtered states.

The rest of the paper is organized as follows. 
In \cref{sec:setup} we introduce the Floquet filter setup and summarize our main results.
In \cref{sec:ee_floquet} we analyze the RDM of Floquet filtered states and compute their R\'enyi-$\alpha$ entanglement entropies. Next we discuss the convergence rate of Floquet filtered states to thermal states in \cref{sec:filter_thermal} and generalize the results to the Hamiltonian regime in \cref{sec:floquet_to_hamil}. Finally, we conclude in \cref{sec:summary_floquet_filter}.

%% file: setup.tex
\section{Setup and main results}
\label{sec:setup}

In this section, we  first introduce the cosine filter~\cite{Ge2019} for local Hamiltonians and discuss how to generalize it to the Floquet framework. Then we  briefly summarize our main results.

\subsection{Cosine filter}
In the Hamiltonian setting, we can reduce the energy variance of a (product) initial state $\ket{\psi}$ by applying an energy filter~\cite{Lu2020} that suppresses spectral weight away from a target value $E$. We define the cosine filter 
\begin{align}
    P_{\delta_H}^{\cos}(H, E):=\cos^{2 M ^2}\left[ \frac{T}{2}(H-E)\right],
    \label{eq:filter_cos}
\end{align}
where $M:=\floor{1 / T \delta_H}$ ($\floor{\cdot}$ being the floor function) and $\sqrt{2}\delta_H$ is the approximate filter width. By writing $\cos x = \left(e^{ix} + e^{-ix}\right)/2$, \cref{eq:filter_cos} can be expanded as
\begin{align}
    P_{\delta_H}^{\cos}(H,E) = \sum_{m=-M^2}^{M^2}c_m e^{-i m T (H-E)},
    \label{eq:filter_cos_sum_exact}
\end{align}
which is a linear combination of evolution operators at discrete times that are (positive and negative) integer multiples of the time step $T$, up to $M^2 T$,  and $c_m$ are the binomial coefficients
\begin{align}
    c_m = \frac{1}{2^{ 2M^2 }}
	\begin{pmatrix}
		2M^2 \\ M^2 - m
	\end{pmatrix}.
    \label{eq:def_cm}
\end{align}

The cosine filter can be approximated by the truncated sum 
\begin{align}
    P_{\delta_H}^{\rm trunc}(H,E) = \sum_{m=-xM}^{xM}c_m e^{-i m T (H-E)},
    \label{eq:filter_cos_sum}
\end{align}
where $x$ is a $\varO(1)$ constant. The cutoff error in the approximation is controlled by $x$ in operator norm and decreases as $2e^{-x^2 / 2}$~\cite{Ge2019}. The maximal evolution time is therefore reduced to $xMT\approx x / \delta_H$. When $x = M$, we will fully recover the untruncated cosine filter.

If the time step $T$ is small enough to ensure that $T(H-E)$ does not take values larger than $2\pi$ on the support of the state, the action of the filter on $\ket{\psi}$ is approximately equal to a Gaussian filter of width $\sqrt{2} \delta_H$~\cite{Ge2019,Lu2020}.
For a local Hamiltonian, this can be ensured choosing $T= \varO(1/N)$, or even $T = \varO(1/\sqrt{N})$ if the state $\ket{\psi}$ is a product state~\citep{Lu2020,Yang2022}. In this case, the energy variance of the resulting filtered state will be approximately $\delta_H^2$ in the small $\delta_H$ limit.

\begin{figure}
	\centering
	\includegraphics[width = .45\textwidth,trim={1.8cm 0.3cm 1.5cm 1cm},clip]{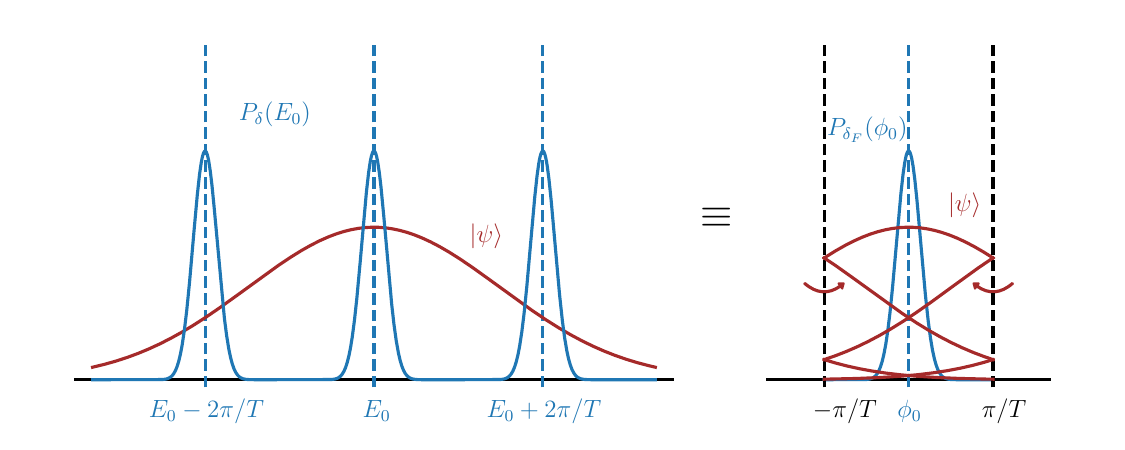}
	\includegraphics[width = .25\textwidth]{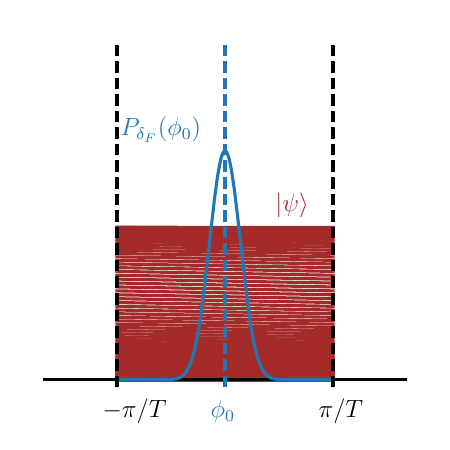}
	\caption{Action of the cosine filtered state $P_{\delta}(E_0)$ on an initial state $\ket{\psi}$ in the Floquet regime. The brown curves represent the local density of states of $\ket{\psi}$, and the blue ones represent the Gaussian filters. \textbf{Upper}: When $2\pi / T$ is smaller than the width of the initial state $\ket{\psi}$, the spectrum will be effectively wrapped around the interval $[-\pi / T, \pi / T)$. \textbf{Lower}: When $T$ is a constant, in the thermodynamic limit the folded spectrum becomes flat and different initial energy scales are sufficiently mixed.}
	\label{fig:filter_hamil_floq}
\end{figure}

\subsection{Floquet filter}
\label{sec:def_floquet_filtering}

The operator in \cref{eq:filter_cos} is periodic, with period $\pi$ in $TH/2$, and thus cannot distinguish eigenvectors whose energies differ in integer multiples of $2 \pi/T$. 
For arbitrary values of $T$, the filter centered at energy $E_0$ is effectively replicated at energies $E_0 + 2\pi n / T$ for all integers $n$, producing a \emph{comb filter} (see \cref{fig:filter_hamil_floq}). 
Alternatively, we can interpret this as a single filter acting on the spectrum folded into the energy window $[-\pi/T, \pi/T)$, which corresponds to the eigenphases of the stroboscopic evolution operator $U(T) = e^{-iHT}$, divided by $T$.
It can thus be seen as a Floquet operator. If $T$ is a constant, in the large $N$ limit $NT \gg 2 \pi$, and the folded spectrum becomes flat. We call this the \emph{fully Floquet regime}.

The truncated Floquet filter can thus be analogously defined as
\begin{equation}\begin{aligned}
    P_{\delta_F}^{\rm trunc}(U, \theta)
    = &   \sum_{m=-xM}^{xM} c_m U^m e^{im\theta}.
\label{eq:floquet_filter}
\end{aligned}\end{equation}
We use the notation $\delta_F:=1/M$ for the Floquet filter width, to be distinguished from the Hamiltonian one $\delta_H$ introduced above. 
Notice that the definition of Floquet filter can be instead taken to extend until $|m|=M^2$, corresponding to the exact expansion of the cosine filter as in \cref{eq:filter_cos_sum_exact}. The resulting filter, which we will denote $P^{\cos}_{\delta_F}(U, \theta)$, is again $e^{-x^2/2}$ close to the truncated Floquet filter $P_{\delta_F}^{\rm trunc} (U, \theta)$ in operator norm. However, as we will argue later, it can result in a different entropy scaling for the filtered states. To distinguish them, we will refer to this extended sum as \emph{cosine} and to the definition in \cref{eq:floquet_filter} as \emph{truncated} Floquet filter. When the distinction is irrelevant, the superscript will be omitted for brevity.

We can write the spectral decomposition of the evolution operator as
\begin{equation}
	U = \sum_k \ket{k}\bra{k}e^{-i \phi_k},
	\label{eq:eigs_u}
\end{equation}
with phases $\phi_k \in[-\pi , \pi )$.
Substituting in \cref{eq:floquet_filter} and comparing to previous subsection, it follows that the Floquet filter approximates a Gaussian of width $\delta_F=\sqrt{2}/M$ centered at $\theta$ in the small $\delta_F$ limit,
\begin{equation}\begin{aligned}
    P_{\delta_F}(U, \theta) \approx
    \sum_k \exp\left[-\frac{(\phi_k - \theta)^2}{4 \delta_F^2} \right] \ket{k}\bra{k}.
\label{eq:floquet_filter_2}
\end{aligned}\end{equation}
Hence the Floquet-filtered state 
\begin{equation}
    \ket{\psi_{\delta_F, \theta}} = \frac{P_{\delta_F} (U, \theta) \ket{\psi}}{\sqrt{ \braket{\psi | P_{\delta_F}^2 (U, \theta) |\psi} }} = \sum_m \tilde{c}_m U^me^{im\theta}\ket{\psi}
    \label{eq:Floquet_filtered}
\end{equation}
will asymptotically have variance $\delta_{F}^2 = 1/ M^2$ in the phase operator $\hat{\phi} = \sum_k \phi_k \ket{k}\bra{k}$. Here $\tilde{c}_m$ are normalized filter coefficients. Their distribution approximates a Gaussian of width $1/2\delta_F$,  schematically represented in \cref{fig:filter_coef} (see \cref{appendix:filter_coef} for details).

\begin{figure}
    \centering 
    \includegraphics[width=.4\textwidth]{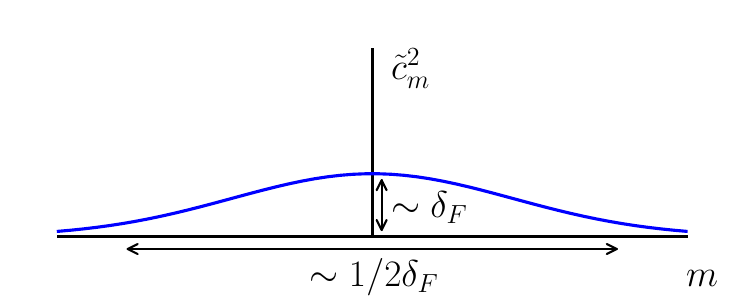}
    \caption{The normalized filter coefficients $\tilde{c}_m$. $\tilde{c}_m$ is approximately a Gaussian function in $m$, whose width is $1 / 2\delta_F$. $\tilde{c}_0^2$ is approximately proportional to $\delta_F$. A similar scaling holds for normalized filter coefficients in the Hamiltonian case (see Appendix~\ref{appendix:filter_coef} for details).}
    \label{fig:filter_coef}
\end{figure}

The eigenphases of the Floquet operator $U = e^{-i HT}$ are $\phi_k = E_k T$ modulo $2\pi$, corresponding to energy eigenvalues $E_k$. We can recover the local Hamiltonian filter when $T = \varO(1 / \sqrt{N})$ if the initial state is a product state, and $T = \varO(1 / N)$ for more general initial states. We call this limit the \emph{Hamiltonian regime}. Instead, for $T = \varO(1)$ we will be in the \emph{fully Floquet regime}. 
In the Hamiltonian regime, the Floquet phase can be interpreted as Hamiltonian energy times $T$. In that case, the energy width of the Hamiltonian filter is related to the Floquet filter phase width by 
\begin{equation}
	\delta_H = 1 / MT = \delta_F / T.
\end{equation}

In the rest of the paper,  unless explicitly indicated, we consider the initial state $\ket{\psi}$ to be a translationally invariant product state, and, for simplicity, the center of the filter to be $\theta = 0$.

\subsection{Main results}
\begin{table*}
	\renewcommand*{\arraystretch}{1.5}
	\centerline{
	\begin{tabular}{|c|c||c|c|c|}
		\hline
		\multicolumn{2}{|c||}{ { \makecell{ \textbf{ R\'enyi-$\alpha$ } \\ \textbf{entanglement entropy} }}}
		& $\alpha > 1$ 
		& $\alpha = 1$ (von Neumann)
		& $0 < \alpha < 1$\\
		\hhline{|==||=|=|=|}
		\multicolumn{2}{|c||}{ \makecell{Hamiltonian\\ (\textcite{Banuls2020}) }}
		&
		& $  \varO\left( \frac{1}{\delta_H}\right) + \log \sqrt{N} $
		& \\
            \hhline{|--||-|-|-|}
		\multicolumn{2}{|c||}{ \makecell{Hamiltonian, 1D worst case \\
        (\textcite{Rai2023}) }} 
		&
		& \makecell{$ \varO\left( \frac{{ \sqrt{\log\frac{N}{\delta_H}} }}{\delta_H}  \right)  + \varO\left( \frac{\log^{\frac{3}{4}} \frac{N}{\delta_H}}{\sqrt{\delta_H}}\right) + \varO\left( \log \frac{N}{\delta_H}  \right)$}
		& \\
            \hhline{|--||-|-|-|}
		\multirow{2}{*}{ \makecell{Hamiltonian\\ (\textcite{Morettini2024}) }  } 
		& $\delta_H = \omega(1)$
		& $\log \frac{\sqrt{N_A}}{\delta_H}$ & $\log \frac{\sqrt{N_A}}{\delta_H}$ 
		& \\
            \hhline{|~-||-|-|-|}
		& $\delta_H = \varO(1)$
		& $\frac{\alpha}{\alpha - 1}\log \frac{1}{\delta_H} + \log \sqrt{{N_A}}$     
		& $ \varO(\frac{1}{\delta_H}) + \log \frac{\sqrt{N_A}}{\delta_H} $
		& $ \varO(\frac{1}{\delta_H^2}) + \log \frac{\sqrt{N_A}}{\delta_H} $\\
            \hhline{|--||-|-|-|}
		\multirow{2}{*}{\centering 
        \makecell{Floquet \\ (This work)}} & \emph{cosine} filter
		& \multirow{2}{*}{\centering $\frac{\alpha}{\alpha - 1}\log \frac{1}{\delta_F}$}
		& \multirow{2}{*}{\centering $\varO({\frac{1}{\delta_F}}) + \log \frac{1}{\delta_F}$}
		& $ \varO({\frac{1}{\delta_F^2}}) + \log\frac{1}{\delta_F} $
		\\
            \hhline{|~-||~|~|-|}
		& Truncated 
		&
		&
		& $ \varO(\frac{x}{\delta_F}) + \log \frac{1}{\delta_F} $\\
		\hline
	\end{tabular}
	}
	\caption{Summary of the scaling of R\'enyi entanglement entropies of filtered states in this and previous works~\citep{Banuls2020,Rai2023,Morettini2024}. 
    We omit $\varO(1)$ terms. $N_A$ is the size of the subsystem. $x$ is the parameter to control truncation errors in the cosine filter. Note that $\delta_F \propto  \delta_H / \sqrt{N}$ in the Hamiltonian limit.
	}
	\label{table:all}
\end{table*}

We consider a system of size $N$, partitioned into subsystems $A$ and $B$ of sizes $N_A$ and $N_B$, respectively,  
both proportional to $N$. For any fixed $\delta_F$, we have $N_A, N_B \gg 1/\delta_F$ in the thermodynamic limit.
In addition, we assume that, for any initial product state $\ket{\psi}$ and any integer $n\neq 0$, the Loschmidt echo $|\braket{\psi | e^{-iHnT} | \psi}|^2$ is exponentially small in the system size $N$. This assumption reflects the expected behavior in the most generic case (see \cref{sec:rdm_floquet} for a more detailed discussion of this condition).
For a fixed Floquet filter width $\delta_F$ and in the thermodynamic limit, we obtain the following results for the filtered state:

\begin{enumerate}
    \item In \cref{sec:rdm_floquet}, we show that the RDM of the subsystem $A$ for the Floquet filtered state $\rho_{A, \delta_F} = \tr_B\left( \ket{\psi_{\delta_F}} \bra{\psi_{\delta_F}}
    \right)$ can be approximately written as the direct sum of RDMs for each step of the Floquet evolution,
    \begin{equation}
        \rho_{A, \delta_F} \sim \oplus_{m} \tilde{c}_m^2 \rho_{A,mm},
    \end{equation}
    where $\rho_{A,mm} = \tr_B\left( U^m \ket{\psi} \bra{\psi}U^{\dagger m} \right)$. In other words, the local Floquet filtered state is equivalent to a weighted time average.

    \item We compute R\'enyi-$\alpha$ entanglement entropies of Floquet filtered states based on the direct sum structure in \cref{sec:mechanism_filter}.
    We find that the entropies exhibit distinct scaling behaviors depending on the order $\alpha$, consistent with the findings previously reported for the Hamiltonian case in~\cite{Morettini2024}.
    \cref{table:all} compiles all entropy scaling results for filtered states.

    \item 
    As the width of the filter decreases, the filtered state approaches thermal properties as $\varO(\sqrt{\delta_F})$. More precisely, the difference in expectation values of any local observable $\hat{O}$ between the filtered state and the thermal one can be bounded as
     \begin{eqnarray}
	   \left| \tr(\rho_{A, \delta_F} \hat{O}) - \tr(\rho_{A,\mathrm{th}}\hat{O}) \right| = \varO\left(\sqrt{\delta_F}\right),
    \end{eqnarray}  
    if assuming Floquet ETH (\cref{sec:filter_thermal}).
\end{enumerate}

In \cref{sec:floquet_to_hamil}, we extend the result to the Hamiltonian regime. Specifically, we show that a Hamiltonian-filtered state $\ket{\psi_{\delta_H}}$ with filter width $\delta_H$ can be approximated by a Floquet-filtered state with filter width $\tilde{\delta}_F = \Theta(\delta_H)$. We finally obtain that:
\begin{enumerate}
\setcounter{enumi}{3}
    \item  Local expectation values in the Hamiltonian-filtered state converge to the thermal ones as~\footnote{We omit logarithmic scalings in system size $N$ with the $\protect\tilde{\varO}$ notation.}
\begin{eqnarray}
	\left|\braket{\psi_{\delta_H} | \hat{O} | \psi_{\delta_H}} -  \tr (\rho_{\rm th}(\psi) \hat{O})\right| = \tilde{\varO}\left(\sqrt{\delta_H} \right).
\end{eqnarray}
\end{enumerate}

%% file: entanglement.tex
\section{Entanglement growth in Floquet filtered state}
\label{sec:ee_floquet}

In this section, we compute the R\'enyi entanglement entropies of the Floquet-filtered states by exploiting the specific structure of their reduced density matrices.

\subsection{The reduced density matrices of Floquet filtered states}
\label{sec:rdm_floquet}
The filtered state can be written as a weighted sum of time evolved states $U^{m}\ket{\psi}$. The overlap between two such terms $U^{m}\ket{\psi}$ and $U^{n}\ket{\psi}$ is related to the Loschmidt echo
\begin{equation}
   L((m-n)T) = |\braket{\psi|U^{m-n}|\psi}|^2.
\end{equation}
The Loschmidt echo typically decays exponentially in system size~\cite{LECLAIR1995581,Heyl2013,Heyl2018}.
In the thermodynamic limit, we can define a Loschmidt rate
\begin{equation}
    f(t) = - \lim_{N\to \infty}  \frac{1}{N}\log_2 L(t).
    \label{eq:ft}
\end{equation}
We assume that for a generic Hamiltonian $f(nT)$ is lower bounded by a positive constant $\gamma_n$ for all initial product states $\ket{\psi}$ and integer $n\neq 0$:
\begin{equation}
    \label{eq:gamma_nT}
    \gamma_n := \inf_{\ket{\psi}} f(nT) > 0, \quad n \neq 0.
\end{equation}
Thus, in the fully Floquet regime, the overlap between states $U^{m}\ket{\psi}$ and $U^{n}\ket{\psi}$ vanishes exponentially in system size if $m\neq n$. This condition reflects the expected behavior of product states for times much shorter than any revival time, in the most generic situation, excluding any symmetries, or product eigenstates. It could fail, for example, in systems with quantum many-body scar states~\cite{Turner2018a, Turner2018, Lin2018a, Khemani2019, Ho2019}, where the Loschmidt echo exhibits significant revivals.

The absence of coherences between different time steps allows us to determine the structure of the RDM of Floquet-filtered states. To be more concrete, let us consider the bipartition of the total system into subsystems $A$ and $B$, with subsystem sizes $N_A, N_B$ proportional to $N$. The initial product state can be correspondingly decomposed as $\ket{\psi} = \ket{\psi_A}\otimes \ket{\psi_B}$. Using \cref{eq:Floquet_filtered} we can write the RDM for $A$ as
\begin{equation}\begin{aligned}
    \rho_{A, \delta_F} = & \sum_{m,n} \tilde{c}_m\tilde{c}_n \tr_B (U^{m}\ket{\psi}\bra{\psi}U^{\dagger n}) \\
    = & \sum_{m,n} \tilde{c}_m\tilde{c}_n \rho_{A, mn},
\end{aligned}\end{equation}
where we have defined $ \rho_{A,mn} =  \tr_B \left(U^{m}\ket{\psi}\bra{\psi}U^{\dagger n} \right)$. 

We observe that for any fixed $\delta_F$, the RDM $\rho_{A, \delta_F}$ of the Floquet filtered state $\ket{\Psi_{\delta_F}}$, generated by a local Hamiltonian with constant $T$, is almost \emph{block-diagonal} with blocks $\rho_{A,mm}$ in the thermodynamic limit. The argument can be sketched as follows
\begin{enumerate}
    \item  The diagonal blocks are non-vanishing. More concretely, we can lower bound their Frobenius norm by a constant.
    
    \item  The diagonal blocks are approximately mutually orthogonal.      
    Different blocks have almost orthogonal supports, i.e. the nonzero eigenvectors of $\rho_{A,mm}$ and $\rho_{A,nn}$  have almost no overlap. 
    We argue that when $m\neq n$,
    \begin{equation}
        \tr \left( \rho_{A,mm} \rho_{A,nn} \right) \simeq e^{-\Omega(N_A)}.
        \label{eq:ortho_diag}
    \end{equation}
    
    \item The off-diagonal blocks approximately vanish, i.e., $\rho_{A,mn} \approx 0$ for any $m \neq n$. To be more precise, the distance (induced by the Frobenius norm) between $\rho_{A,\delta_F}$ and the sum of diagonal blocks is exponentially small in $N_B$:
    \begin{align}
        \left\|\rho_{A, \delta_F} - \sum_{m} \tilde{c}_m^2  \rho_{A, mm}\right\|_F^2 \simeq e^{-\Omega(N_B)}.
        \label{eq:condition3}
    \end{align}
\end{enumerate}

\begin{figure}
    \centering
    \includegraphics[width=0.45\textwidth]{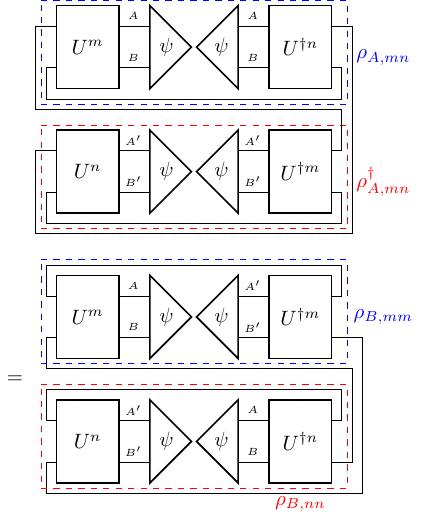}
    \caption{Diagram for \cref{eq:otho_diag0}. We swap the upper and lower blocks in the right hand side.}
    \label{fig:circuit_contration}
\end{figure}

\paragraph*{Sketch of the argument.-}
Condition 1 is straightforward. In our setup, we fix $\delta_F$, and thus $m,n \le x / \delta_F$ are both upper bounded by a constant. Since the R\'enyi-2 entanglement entropy of local interacting systems grows at most linearly in time~\cite{Shi2024,Toniolo2025}: $S_2(t) \le Ct$ for some constant $C$, the purity of the diagonal density matrices is lower bounded by a constant
    \begin{equation}
        \tr \left( \rho_{mm}^2 \right) = 2^{-S_2(mT)} \ge 2^{-xCT / \delta_F}.
    \end{equation}

Conditions 2 and 3 are indeed related to each other. The Frobenius norm of an off-diagonal block $\rho_{A, mn}$ can be rewritten as the overlap of two diagonal blocks in the complementary subsystem,
\begin{equation}\begin{aligned}
    \norm{\rho_{A, mn}}_F^2 & = \tr \left( \rho^{\dagger}_{A, mn} \rho_{A, mn} \right) = \tr\left( \rho_{B, mm} \rho_{B, nn} \right),
\label{eq:otho_diag0}
\end{aligned}\end{equation}
where $\rho_{B, mm} = \tr_A \left( U^{m}\ket{\psi} \bra{\psi} U^{\dagger m} \right)$ is the reduced density matrix of the states evolved for $m$ stroboscopic time steps. The equation is visualized in the diagram in \cref{fig:circuit_contration}, which comes from swapping the upper and lower blocks in the right hand side.
Therefore, we only need to show that the overlap $\tr \left( \rho_{A,mm} \rho_{A,nn} \right)$ is exponentially small in $N_A$ if $m\neq n$, or, analogously $\tr \left( \rho_{B,mm} \rho_{B,nn} \right)$ is exponentially small in $N_B$, and the bound \eqref{eq:condition3} for the off-diagonal contributions will follow.

The special case $m=0$ has been studied previously in several numerical~\cite{Bandyopadhyay2021,Halimeh2021} and experimental~\cite{karch2025} works. Intuitively, we expect the same property to hold for generic cases with $m,n\neq 0$. By the Lieb-Robinson bound, when $m$ and $n$ are at most of order $1/\delta_F$, the relevant finite-time evolutions approximately factorize between the subsystems $A$ and $B$, i.e., $U^k \approx U_A^k\otimes U_B^k$ for such $k$. Consequently, $\tr(\rho_{A,mm}\rho_{A,nn})$ is expected to scale as a subsystem Loschmidt echo. This heuristic argument can be made rigorous using the result of Ref.~\cite{Osborne2006Lieb}, with an additional correction supported only near the boundary between $A$ and $B$. Assuming the generic exponential decay of product-state Loschmidt echoes with system size, we then obtain the desired condition \eqref{eq:ortho_diag}. The detailed argument is given in \cref{appendix:proof_ortho}.

Because the same argument can be applied indistinctly to subsystems $A$ or $B$, using \eqref{eq:otho_diag0} we conclude that $\norm{\rho_{A, mn}}_F^2 \simeq e^{-\Omega(N_B)}$. This in turn can be used to bound the Frobenius norm of the sum of off-diagonal blocks from \eqref{eq:condition3} as 
\begin{align}
    \left\|\sum_{m\neq n} \tilde{c}_m \tilde{c}_n   \rho_{A, mn}\right\|_F & \leq
    \sum_{m\neq n} |\tilde{c}_m \tilde{c}_n | \norm{\rho_{A, mn}}_F  \nn \\
    & \leq
    (xM)^2 \norm{\rho_{A, mn}}_F  
    \simeq e^{-\Omega(N_B)}.
\end{align}

\begin{figure}
    \centering
    \includegraphics[width = .225\textwidth]{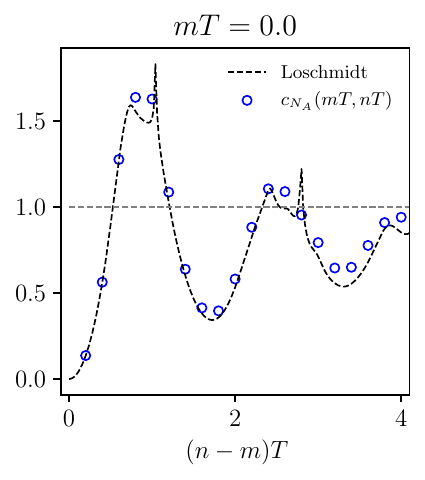}
    \includegraphics[width = .225\textwidth]{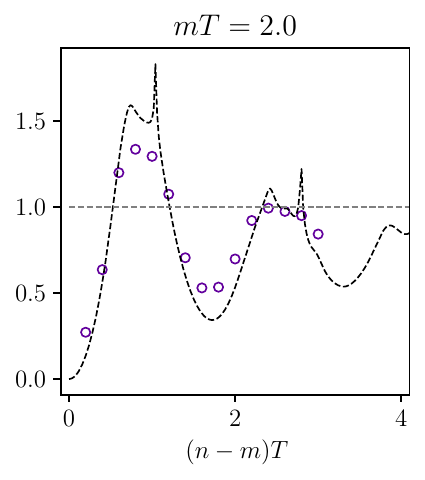}
    \caption{Overlap between diagonal blocks $m$ and $n$ of the RDM as a function of the time difference, for the non-integrable Ising model defined in \cref{eq:Hising}. The system size $N=20$, subsystem size $N_A = 10$, and $T=0.2$.
    Circles represent the correlation rate  in \cref{eq:def_correlation}, defined as the logarithm of the overlap rescaled by the subsystem size. The black dashed line corresponds to $f(t)$, the return rate in \cref{eq:ft}, computed for $N=20$. Numerical data were obtained via matrix product state (MPS) simulations.
    }
    \label{fig:rdmoverlap}
\end{figure}

\paragraph*{Numerical verification.-}
To numerically test this scaling, we compute the overlap $\tr \left( \rho_{A,mm} \rho_{A,nn} \right)$ for the mixed field Ising model
\begin{equation}
    H = J\sum_{i = 1}^{N-1} \sigma^z_i \sigma^z_{i+1} + g\sum_i \sigma^x_i + h \sum_i \sigma^z_i,
    \label{eq:Hising}
\end{equation}
with $(J, g, h) = (1, -1.05, 0.5)$. We choose different $0\le m<n$, and total system size $N=20$. The subsystem size is $N_A=10$. The results are shown in~\cref{fig:rdmoverlap}, where the $y$-axis corresponds to
\begin{equation}
    c_{N_A}(mT,nT):=-\frac{1}{N_A}\log_2 \tr \left( \rho_{A,mm} \rho_{A,nn}\right).
    \label{eq:def_correlation}
\end{equation}
We observe that $c_{N_A}(mT, nT)$ (circles) is very close to $f((m-n)T)$ (black dashed lines), which coincides with the subsystem Loschmidt echo argument.

We have verified all the block-diagonal conditions of the RDM $\rho_{A, \delta_F}$.
These conditions indicate that the RDM  behaves like the weighted direct sum of the diagonal blocks $\rho_{A,mm}$
\begin{equation}
    \rho_{A, \delta_F} \sim \oplus_{m} \tilde{c}_m^2 \rho_{A,mm}.
    \label{eq:blocks}
\end{equation}
Notice that the matrix on the right hand side has in general a much larger dimension (up to $O(M \mathrm{dim}(A))$), and the equivalence holds when the respective supports of the left and right expressions are considered, 
meaning that eigenvalues of $\rho_{A, \delta_F}$ can be obtained from the set of non vanishing eigenvalues of each $\rho_{A, mm}$ rescaled by $\tilde{c}_m^2$, with corrections up to order $e^{-\Omega(N_A)}$ (see \cref{appendix:eigvals} for details). This effective eigenvalue distribution can be employed to approximate the Rényi entanglement entropies (also see  \cref{appendix:eigvals}), as we do in the following subsection. Additionally, \cref{appendix:floquet_toy} describes in detail an exactly solvable Floquet toy model where this direct sum structure is exact.

\subsection{R\'enyi entanglement entropy of local Floquet models}
\label{sec:mechanism_filter}

Let us denote the nonzero eigenvalues of the block $\rho_{A, mm}$ by $\{\sigma_{m,k}\}_k$.
Using the block structure of the RDM of Floquet-filtered states in \cref{eq:blocks}, the nonzero eigenvalues of $\rho_{A,\delta}$ can be estimated by $\left\{ \tilde{c}_m^2 \sigma_{m,k} \right\}_{m,k}$. We can now  
evaluate the Rényi-$\alpha$ entanglement entropies for different orders $\alpha$, with two additional ingredients:
\begin{itemize}
    \item For large $M$, or narrow enough filters, the normalized coefficient $\tilde{c}_m^2$ scales with $m$ as a Gaussian centered at 0 with width $1 / 2\delta_F = M / 2$ (see \cref{appendix:filter_coef}):
    \begin{align}
        \label{eq:filter_coef}
        \tilde{c}_m^2 \propto \exp\left(-2m^2/M^2\right)/M.
    \end{align} 
    \item We assume that $S_{\alpha}(\rho_{A,mm}) = k_{\alpha} |m|T$ grows linearly in $m$ for any $\alpha$, which is the typical case for generic local Hamiltonians~\cite{Calabrese2005b,Bravyi2007,Fagotti2008,Kim2013a,Bertini2022,Toniolo2025}.
\end{itemize}

\paragraph{Von Neumann entanglement entropy.}
For $\alpha=1$, we can write
\begin{align}
    S_1(\rho_{A, \delta_F}) \approx & - \sum_{m, k} \tilde{c}_m^2 \sigma_{m,k} \log (\tilde{c}_m^2\sigma_{m,k}) \nn \\
    = & \sum_{m} \tilde{c}_m^2 S_1(\rho_{A, mm}) + S_1(\left\{ \tilde{c}_m^2 \right\}),
\label{eq:renyi_ent_eq1}
\end{align}
where $S_1(\rho_{A, mm})=\sum_{k} \sigma_{m,k} \log \sigma_{m,k}$ are the entanglement entropies of each evolved state $U^m\ket{\psi}$,
and $S_1(\left\{ \tilde{c}_m^2 \right\})=\sum_{m} \tilde{c}_m^2 \log \tilde{c}_m^2$ is the Shannon entropy of the coefficients $\tilde{c}_m^2$. When $S_1(\rho_{A,mm}) = k_1 |m| T$ grows linearly in $m$, the first term will scale as $1 / \delta_F$ as well. The second term can be directly computed by inserting the Gaussian function, and turns out to be logarithmic in $1 / \delta_F$. Adding them together, we have
\begin{equation}
	S_1(\rho_{A, \delta_F}) = \frac{k_1 T}{\sqrt{2\pi} \delta_F} + \log \frac{1}{\delta_F}  + \varO(1).
	\label{eq:floquet_vN_entropy}
\end{equation}

\paragraph{R\'enyi-$\alpha$ entanglement entropy with $\alpha\neq 1$.}
When $\alpha \neq 1$, the R\'enyi-$\alpha$ entanglement entropy can be analogously written as 
\begin{equation}\begin{aligned}
    S_{\alpha} (\rho_{A,\delta_F}) & \approx \frac{1}{1 - \alpha} \log \left[ \sum_{m, k } (\tilde{c}_m^2 \sigma_{m,k})^{\alpha}  \right]\\
    & = \frac{1}{1-\alpha} \log \left[ \sum_{m} \tilde{c}_m^{2\alpha} 2^{(1 - \alpha) S_{\alpha}(\rho_{A, mm})} \right].
    \label{eq:renyi-weighted}
\end{aligned}\end{equation}
In this case, the $m=0$ term (corresponding to the initial state) contributes a single large eigenvalue $\tilde{c}_0^2\propto 1/ M = \delta_F$ to the reduced density matrix. When $\alpha > 1$, this term dominates the sum, and gives a bound
\begin{equation}
	S_{\alpha > 1}(\rho_{A,\delta_F}) < \frac{\alpha}{1-\alpha} \log {\tilde{c}_0^2} \sim  \frac{\alpha}{\alpha-1} \log \frac{1}{\delta_F}.
	\label{eq:renyi_ent_gtr1}
\end{equation}

For $0<\alpha<1$, we proceed by bounding the sum in \cref{eq:renyi-weighted} by the number of terms [$\varO(2/\delta_F)$] times the largest one, so that 
\begin{align}
& (1-\alpha)S_{\alpha < 1}(\rho_{A,\delta_F}) \nn \\
\leq & \max_m \left \{ 
 \alpha \log \tilde{c}_m^2  + (1 - \alpha) S_{\alpha}(\rho_{A,mm}) \right\} 
 + \log(2 / \delta_F)  \nn  \\
= & \max_m \left \{ - 2  \alpha m^2 \delta_F^2 + (1 - \alpha) k_{\alpha} |m| T \right\}   + \varO(\log(1 / \delta_F)).
    \label{eq:max_xm}
\end{align}

\cref{eq:max_xm} is a quadratic function in $m$ when $m > 0$, whose maximum value is taken at $m = m_0 := (1 - \alpha) k_{\alpha}T / 4 \alpha \delta_F^2$, and monotonically increasing when $0 < m < m_0$. For the truncated filter defined in \cref{eq:floquet_filter}, $m \le x/\delta_F$, and thus the maximal value of \cref{eq:max_xm} is attained at $m=x/\delta_F$ for narrow filters with large $1 / \delta_F$. Therefore
\begin{equation}
	S_{\alpha < 1}(\rho_{A,\delta_F}^{\rm trunc}) \sim x \cdot \frac{k_{\alpha} T}{\delta_F} +  \varO \left(\log \frac{1}{\delta_F}\right)
	\label{eq:renyi_ent_less1p}
\end{equation}
is linear in $1 / \delta_F$.

It is interesting to notice that this scaling changes for the untruncated cosine filter $P_{\delta_F}^{\cos}$, in which the limit of the sum in \cref{eq:floquet_filter} is taken to be $1 / \delta_F^2$, as it would correspond to the exact cosine filter~\cite{Ge2019,Lu2020}. In such case, the maximum of \cref{eq:max_xm}
is reached at 
$m = \min\left\{ m_0, 1 / \delta_F^2 \right\}$, and thus
\begin{equation}
	S_{\alpha<1}(\rho_{A,\delta_F}^{\mathrm{cos}}) = \varO\left(\frac{T}{\delta_F^2}\right) + \varO\left(\log \frac{1}{\delta_F}\right).
	\label{eq:renyi_ent_less1}
\end{equation}
If $ \alpha > 1 / (4 + k_\alpha T)$, the maximal value is reached at $m = m_0$, and the coefficient of the quadratic term is ${ (1 - \alpha)  k_{\alpha}^2 T^2 } / 8 \alpha$, coinciding with the results  for the Hamiltonian case in \cite{Morettini2024}.

We further validate our results in \cref{appendix:solvable_eg} using two exactly solvable benchmarks: a toy model mimicking free-particle propagation and a global random unitary model. Due to their technical complexity, the detailed derivations for these models are deferred to the Appendix.

Our findings, together with the more recent analysis of \citet{Morettini2024} in the Hamiltonian regime\footnote{\citet{Morettini2024} employ the replica trick to compute Rényi entanglement entropies under two assumptions: linear growth of entanglement entropy for time-evolved states, and Gaussian decay of Loschmidt-echo-like quantities.}, suggest a universal logarithmic scaling in $1/\delta_F$ (or $1/\delta_H$) when $\alpha > 1$. For $\alpha = 1$, aside from the worst-case scenario, the entanglement entropy grows similarly to that of the initial state evolved up to time $\varO(1/\delta_F)$ (or $\varO(1/\delta_H)$), albeit with a $\sqrt{2\pi}$ times slower growth rate as shown in \cref{eq:floquet_vN_entropy}. In contrast, for $\alpha<1$, $S_{\alpha < 1}$ can increase even more rapidly than under direct time evolution.

%% file: thermality.tex
\section{How thermal is a Floquet filtered state?}
\label{sec:filter_thermal}

We now return to our original question: how thermal is a filtered pure state? In the generic case, a thermal pure state~\cite{Sugiura2012}, i.e. a state that locally resembles the microcanonical ensemble,
should exhibit volume-law Rényi-$\alpha$ entanglement entropy for any $\alpha > 0$, since the density of states at finite energy densities grows exponentially with system size. Nevertheless, we have shown that for Floquet filters generated by local Hamiltonians with $1/\delta_F = \mathrm{poly}(N)$, the Rényi-$\alpha$ entanglement entropies scale only logarithmically with $N$ when $\alpha > 1$. This demonstrates that filtered states cannot be fully thermal when $1/\delta_F$ scales polynomially in system size.

However, in \cref{sec:mechanism_filter}~\cref{eq:condition3} we argued that the RDM of the filtered state can be approximated by the sum of its diagonal terms, with exponentially small Frobenius norm distance error in $N_B$. More concretely, the explicit bound obtained in\cref{appendix:proof_ortho} reads
\begin{equation}
    \tr \left( \rho_{B,mm}\rho_{B,nn} \right) \le \alpha_{\delta_F} 2^{- \beta_{\delta_F} N_B}
\end{equation}
with constants $\alpha_{\delta_F}$ and $\beta_{\delta_F}$ depending only on $\delta_F$. For any local observable $\hat{O}$ in subsystem $A$ with normalization condition $\norm{O} = 1$, we have $\norm{\hat{O}}_F^2 \le 2^{N_A}$, and thus
\begin{equation}
    \begin{aligned}
        & \left|\tr \left[ \left(\rho_{A,\delta_F} - \sum_{m}\tilde{c}_m^2 \rho_{A,mm} \right) \hat{O} \right]\right| \\
        \le &  \norm{\rho_{A,\delta_F} - \sum_{m}\tilde{c}_m^2 \rho_{A,mm}}_F \norm{\hat{O}}_F\\
        \le & (xM)^2\sqrt{\alpha_{\delta_F}} 2^{- (\beta_{\delta_F} N_B - N_A ) / 2}.
    \end{aligned}
\end{equation}

Given $\delta_F$, we may thus choose $N_A \le \beta_{\delta_F}N_B/2$. This ensures that the error
$
\left|\tr \left[ \left(\rho_{A,\delta_F}-\sum_m \tilde c_m^2 \rho_{A,mm}\right)\hat O \right]\right|
= e^{-\Omega(N)}$
vanishes in the thermodynamic limit.
In this sense the filtered state is equivalent to the time average weighted by $\tilde{c}_m^2$:
\begin{equation}\begin{aligned}
    \braket{\psi_{\delta_F} | \hat{O}  | \psi_{\delta_F}} \approx
    &\sum_m \tilde{c}_m^2 \tr \left( \rho_{A,mm} \hat{O} \right)\\
    = & \sum_{m} \tilde{c}_m^2 \braket{ \psi | U^{\dagger m} \hat{O} U^m |\psi}\\
    = & \sum_{k,k'} \psi_{k}^{*} \psi_{k'} \sum_{m} \tilde{c}_m^2 e^{im(\phi_k - \phi_{k'})}  O_{k,k'} \\
    \approx & \sum_{k,k'} \psi_{k}^{*} \psi_{k'} e^{-\frac{(\phi_k - \phi_{k'})^2}{8\delta_F^2} } O_{k,k'}
\end{aligned}
\label{eq:time_average_filter}
\end{equation}
where $U = \sum_k \ket{k}\bra{k} e^{-i \phi_k}$ is the spectral decomposition of $U$ as in \cref{eq:eigs_u}, $\psi_k = \braket{k | \psi}$, and $O_{k,k'} = \braket{k | \hat{O} | k'}$. We have used that $\tilde{c}_m^2$ is approximately a Gaussian function with width $1 / 2\delta_F$ and centered at 0 as in \cref{eq:filter_coef}. The last step corresponds to its Fourier transform. \cref{eq:time_average_filter} shows that the contributions from the off-diagonal elements $O_{k,k'}$ with $k \neq k'$ are suppressed by a Gaussian. In the rest of this section, we analyze how the contribution of these off-diagonal terms depends on the filter width.

In generic systems, the matrix elements $O_{k,k'}$ of a physical observable in the energy eigenbasis are expected to satisfy the ETH ansatz.
In the case of Floquet systems satisfying the corresponding Floquet ETH, observable matrix elements fulfill the random matrix theory properties~\citep{DAlessio2014}:
\begin{equation}
	O_{k,k'} = \bar{O} \delta_{k,k'} + \sqrt{\frac{\overline{O^2}}{\mathcal{D}} } R_{k,k'},
	\label{eq:eth2}
\end{equation}
where $\mathcal{D}$ is the Hilbert space dimension, $\bar{O} = \tr O / \mathcal{D}$, $\overline{O^2} = \tr (O^2) / \mathcal{D}$ (below we assume the normalization condition $\overline{O^2} = 1$), and $R_{k,k'}$ are Gaussian-distributed complex random variables with zero mean and unit variance.

Under this assumption, the deviation of the expectation value in \cref{eq:time_average_filter} with respect to the thermal value $\bar{O}$ is given solely by the fluctuations in the second term of \cref{eq:eth2}, and can be written as the expectation value of a matrix, 
\begin{align}
    & \braket{\psi_{\delta_F} | \hat{O}  | \psi_{\delta_F}}-\mathrm{tr}(\rho_{\mathrm{th}}\hat{O}) \nn \\
     \approx & \sum_{k,k'}\psi_k^{*} \psi_{k'} e^{-\frac{(\phi_k - \phi_{k'})^2}{8\delta_F^2} } R_{k,k'} / \sqrt{\mathcal{D}} \nn \\
     =& \braket{\psi | Q(\delta_F) | \psi},
     \label{eq:deviation}
\end{align}
where $\rho_{\mathrm{th}}$ refers to the diagonal ensemble of the initial state $\ket{\psi}$, and we have defined the matrix $Q(\delta_F)$ with components in the energy eigenbasis
\begin{eqnarray}
	Q_{k,k'}(\delta_F) = R_{k,k'} e^{-\frac{(\phi_k - \phi_{k'})^2}{8\delta_F^2}} / \sqrt{\mathcal{D}}.
\end{eqnarray}
Since \cref{eq:deviation} is bounded by the largest eigenvalue of $Q(\delta_F)$, we aim to bound the spectral radius of this matrix. To be more specific, let us denote the eigenvalues of $Q(\delta_F)$ by $\lambda_p(\delta_F)$ with $1 \le p \le \mathcal{D}$. Our goal is to show that 
\begin{equation}
    \lambda_p(\delta_F) = \varO\left(\sqrt{\delta_F}\right).
\end{equation}

The simplest approach to treat this problem is to assume that all the random variables $R_{k,k'}$ are independent. In this case, the matrix $Q\sqrt{\mathcal{D}}$ is similar to a random band matrix whose band width proportional to $\delta_F {\mathcal{D}}$. The eigenvalues of $Q\sqrt{\mathcal{D}}$ will follow the semi-circle law with radius $\propto \sqrt{\delta_F {\mathcal{D}}}$~\cite{Kus1991,Molchanov1992}, and thus $\lambda_p(\delta_F) = \varO\left(\sqrt{\delta_F}\right)$ directly follows.
A similar argument was used in \cite{Dymarsky2019} to discuss the relation between canonical universality and ETH. 
But already in that work it was pointed out that the independence among $R_{k,k'}$ is not justified from fundamental arguments~\cite{Dymarsky2022}. Indeed, this condition would mean that all four-point or higher order correlators can be computed through the Wick's theorem from two-point correlators, indicating that all thermal states are Gaussian.

The order of such correlations can however be estimated through their contributions to higher order correlators of bounded observables. Recent works~\cite{Foini2019,Foini2019a} suggest that, to ensure $\braket{ \hat{O}^n } = \varO(1)$ for a bounded observable $\hat{O}$, the expectation values of $n$-th order correlations among $R_{k,k'}$ should have the form 
\begin{eqnarray}
    \overline{ R_{k_1,k_2} R_{k_2,k_3}\cdots R_{k_n,k_1} } = F_n / \mathcal{D}^{\frac{n}{2}-1}
\end{eqnarray}
when all the indices $k_i$ ($i=1,\ldots n$) are different, and the coefficient $F_n = \varO(1)$.

We can estimate the eigenvalues of $Q(\delta_F)$ taking into account this generalized ETH ansatz~\cite{Foini2019,Murthy2019}.
To be more specific, we compute the moments $\tr \left[ Q(\delta_F)^{2l} \right]$ for $l\in \mathbb{N}^{+}$ and show that
\begin{equation}
    \tr \left[ Q(\delta_F)^{2l} \right] = \sum_{p} \lambda_p^{2l} (\delta_F) = \varO(\mathcal{D}\delta_F^l),
    \label{eq:moments}
\end{equation}
In \cref{eq:moments}, the higher order correlations among $R_{k,k'}$ only contribute to a higher order correction in $\delta_F$, and thus vanish in the small $\delta_F$ limit. The details are deferred to \cref{appendix:q_delta}. 
Therefore $\lambda_p(\delta_F) = \varO\left( \sqrt{\delta_F} \right)$, and
\begin{eqnarray}
	| \tr(\rho_{A, \delta} \hat{O}) - \tr(\rho_{A,\mathrm{th}}\hat{O}) | = \varO\left(\sqrt{\delta_F}\right).
\end{eqnarray}

Notice that our argument corroborates previous numerical observations obtained for Hamiltonians. In~\cite{Dymarsky2019}, the canonical universality hypothesis was introduced for chaotic systems, stating that 
when a state is supported in a narrow energy interval of width $\delta$, the difference in expectation values of observables with respect to thermal ones is typically bounded by 
$\varO(\delta^{1/\eta})$.
A value $\eta\approx 2$ was found from numerical simulations. The canonical universality in~\cite{Dymarsky2019} was derived from ETH plus the additional assumption of independently distributed $R_{k,k'}$, which is not implied in standard ETH, but can be considered a complementary property.
Our result shows that, for Floquet filtered states, canonical universality does directly follow from Floquet ETH, even if the correlations among off-diagonal matrix elements of observables are taken into account.

Another related numerical result was obtained in~\cite{Cakan2021}, where the density operator of the initial state was filtered to suppress off-diagonal elements in the energy eigenbasis, using a Gaussian filter of the Hamiltonian commutator on the vectorized operator. For local observables on subsystem $A$, this is equivalent to \cref{eq:time_average_filter}.  
The numerical evidence in \cite{Cakan2021} showed that the difference between local observable expectation values in the diagonal ensemble and the density matrix defined in \cref{eq:time_average_filter} is bounded by $\delta^{p}$, where $p$ varies from $0.52$ to $0.62$ for different initial states and observables. These results support our arguments, and indicate that they can be extended to the Hamiltonian regime. In the next section, we  show that this is indeed the case.

\section{Application to filtered states in the Hamiltonian regime}
\label{sec:floquet_to_hamil}

The arguments in the Floquet regime can be adapted with only minor modifications to the Hamiltonian regime, allowing us to address the question of thermality also for Hamiltonian-filtered states. As already discussed in \cref{sec:def_floquet_filtering}, for an initial product state $\ket{\psi}$, whose width in energy with respect to a local Hamiltonian is generically proportional to $\sqrt{N}$, we can choose $T  \propto 1 / \sqrt{N}$ to be in the Hamiltonian regime. With this choice of $T$, however, the Loschmidt echos at stroboscopic times $mT$ no longer vanish in the thermodynamic limit, and thus the previous results do not directly apply. 
We can resolve this issue by mapping the setup to an effective Floquet-filtered state.

When $\delta_H = o(1)$, i.e., $M = 1 / \delta_H T =  \omega(\sqrt{N})$, the individual time evolution terms in a filtered state can be grouped into sets of $n$ consecutive terms, where $n \propto \sqrt{N}$ is chosen to be an even integer:
\begin{equation}\begin{aligned}
    &\ket{\psi_{\delta_H}} = \sum_{m = -xM}^{xM} \tilde{c}_m U^{m}\ket{\psi} \\
    \approx & \sum_{k = -\floor{xM / n - 1 / 2}}^{\floor{xM / n - 1 / 2} } U^{k n} \left(  \sum_{\ell = -n / 2 + 1}^{n / 2} \tilde{c}_{k n + \ell} U^\ell \ket{\psi} \right).
\end{aligned}\end{equation}
We drop the terms in the tails that are not grouped, which effectively amounts to slightly changing the cutoff coefficient $x$ in \cref{eq:floquet_filter}. Within each group, the difference in the evolution time between the first and last terms is $nT = \Theta(1)$. Since the normalized filter coefficients $\tilde{c}_m$ are approximately a smooth Gaussian function in $mT$ with width $1 / \delta_H$, inside each group the difference between coefficients can be bounded by $\varO(n / M^{3/2})$ (see \cref{appendix:diff_states} for details).
Defining $\ket{\Psi_n} = \sum_{\ell = -n / 2 + 1}^{n / 2} U^\ell \ket{\psi}$, the Hamiltonian filtered state can be approximated as 
\begin{eqnarray}
	\ket{\psi_{\delta_H}} \approx  \ket{\tilde{\psi}_{\delta_H}} = \sum_{k = - \floor{xM / n - 1 / 2}}^{\floor{xM / n - 1 / 2}} \tilde{c}_{kn} U^{k n}  \ket{\Psi_n}.
	\label{eq:effect_filter}
\end{eqnarray}
We can show that the trace distance $d$ between $\ket{\psi_{\delta_H}}$ and the normalized approximation $\ket{\tilde{\psi}_{\delta_H}} / \tilde{\mathcal{N}}$ is bounded by (see \cref{appendix:diff_states} for details)
\begin{equation}\begin{aligned}
    d & = \sqrt{1 - |\braket{ \psi_{\delta_H} | \tilde{\psi}_{\delta_H} }|^2  / \tilde{\mathcal{N}}^2 } = \tilde{\varO}(\sqrt{\delta_H} ).
\end{aligned}\end{equation}

In the approximated state $\ket{\tilde{\psi}_{\delta_H}}$, $\ket{\Psi_n}$ can be viewed as the (unnormalized) \emph{effective initial state} and $U^n$ is the \emph{effective Floquet operator}. Since the effective time step is $\tilde{T} = nT = \Theta(1)$, it lives in the fully Floquet regime and the diagonality conditions of the RDM are again valid as in \cref{eq:otho_diag0}. Since the maximal evolution time in the truncated Hamiltonian cosine filter $x / \delta_H$ and that in the Floquet filter $\tilde{T} \cdot x / \tilde{\delta}_F$ are equal to each other, the effective Floquet filter width is $\tilde{\delta}_F = \tilde{T} \delta_H = \Theta(\delta_H)$. The filtered state in the Hamiltonian regime can thus be reduced to one in the Floquet regime. 

Note that compared with  \cref{sec:mechanism_filter}, the effective initial state $\ket{\Psi_n}$ is no longer a product state. It is instead a \emph{pre-filtered} state whose width is only of order $\varO(1)$, which is the reason why the effective Floquet filter will not cause folding of the spectrum. 
The conditions in \cref{sec:rdm_floquet} still hold, which can be seen by expanding $\ket{\Psi_n}$ according to its definition.
In the Hamiltonian case, the ETH modifies \cref{eq:eth2} with smooth functions of the eigenvalues of $k$ and $k'$ that modulate the diagonal and off-diagonal terms (see~\cite{DAlessio2014}). 
Since we focus on an energy interval of constant width, this functions can be considered constant (as they vary over extensive range of energies) and the same form as \cref{eq:eth2} can be used to determine the scaling of the off-diagonal matrix elements $O_{k,k'}$. 
The same analysis thus applies in bounding the errors from thermal states, and we finally obtain
\begin{align}
	|\braket{\psi_{\delta_H} | \hat{O} | \psi_{\delta_H}} -  \tr (\rho_{\rm th}(\psi) \hat{O})| = \tilde{\varO}\left(\sqrt{\delta_H}  \right).
    \label{eq:obs_Hamil}
\end{align}

\begin{figure}
	\centering
	\includegraphics[width = .45\textwidth]{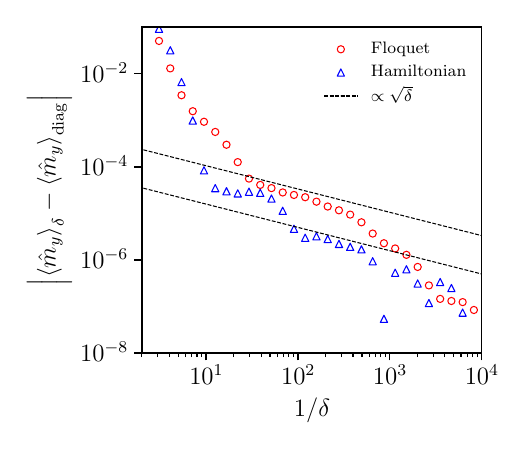}
	\caption{The difference between filtered state and diagonal ensemble expectation values of observable $\hat{m}_y = \sum_{i = 1}^{N} \sigma^y_i / N$ in the nonintegrable Ising chain. The initial state is chosen to be $\ket{Y+} = \left(\frac{1}{\sqrt{2}}(\ket{0} + i \ket{1}) \right)^{\otimes N}$. The system size $N = 15$ and the calculations are performed by exact diagonalization. In the Floquet case, the stroboscopic time step is chosen to be $T = 1$. After the initial fast relaxation, for small $\delta$, in both Floquet and Hamiltonian cases the difference between filtered and thermal values are bounded by $\varO(\sqrt{N})$.}
	\label{fig:diff_thermal_filter}
\end{figure}

We verify this scaling numerically in \cref{fig:diff_thermal_filter}, where we show exact diagonalization results for the non-integrable Ising chain with system size $N = 15$.  
The initial state in these simulations is chosen to be $\ket{Y+} = \left[\frac{1}{\sqrt{2}}(\ket{0} + i \ket{1}) \right]^{\otimes N}$, and the observable is $\hat{m}_y = \sum_{i = 1}^{N} \sigma^y_i / N$. This choice guarantees the initial expectation value $\langle Y+ | \hat{m}_y | Y+ \rangle = 1$, which is far away from the diagonal ensemble  value $\langle \hat{m}_y\rangle_{\mathrm{diag}}=0$. The plot shows how, after the fast initial relaxation,
the convergence rate is bounded by $\varO(\sqrt{\delta})$ for both Floquet and Hamiltonian case. 

Unlike local observable expectation values, the entanglement entropy results are not as straightforward to generalize from the Floquet regime to the Hamiltonian regime.
While we can compute those of the Floquet approximated state $\ket{\tilde{\psi}_{\delta_H}}$ following the same steps as \cref{sec:mechanism_filter}, they are not theoretically guaranteed to share the same scaling of the entanglement entropies of $\ket{\psi_{\delta_H}}$. Nevertheless, the entanglement entropies of the $\ket{\tilde{\psi}_{\delta_H}}$ are indeed in line with the results obtained in \cite{Morettini2024} (see \cref{table:all}). 

While in previous settings the initial state is a product state, and hence no entanglement, here the pre-filtered effective initial state $\ket{\Psi_n}$ has entanglement entropy bounded by $\varO(\log \sqrt{N})$, which adds an additional contribution to all R\'enyi entanglement entropies. This find this bound from the bond dimension of the MPS representation of $\ket{\Psi_n}$, as follows. Each term $U^m \ket{\psi}$, i.e. the initial product state $\ket{\psi}$ evolved to time $mT \propto m / \sqrt{N}$, can be faithfully represented by an MPS of bond dimension $D_0^{|m| / \sqrt{N}}$ with some constant $D_0$. Hence, the weighted sum of such terms in $\ket{\Psi_n}$ requires a bond dimension at most 
\begin{align}
    D_n \le & \sum_{m=-n/2+1}^{n/2} D_0^{m / \sqrt{N}} 
    \leq \frac{2 - D_0^{\frac{n/2+1}{\sqrt{N}}} - D_0^{\frac{n/2}{\sqrt{N}}} }{1 - D_0^{\frac{1}{\sqrt{N}}}}
    \nn \\
    \approx & \frac{2(D_0^{c}-1)}{\ln D_0}\sqrt{N},
\end{align}
for a constant $c$.
Therefore R\'enyi entanglement entropies of $\ket{\Psi_n}$ are bounded by $\varO(\log \sqrt{N})$ for any $\alpha > 0$. This initial state contribution should be included when calculating the entanglement entropy for a Hamiltonian-filtered state. For instance, the von Neumann entanglement entropy is:
\begin{equation}
	\label{eq:entropy_hamil}
	S_1(\rho_{A, \delta_H}) = \log \frac{\sqrt{N}}{\delta_H} + \frac{k_1}{\sqrt{2\pi} \delta_H} + \varO(1).
\end{equation}

If we consider the same maximal evolution time in the truncated cosine filter, i.e. $x / \delta_H$ in the Hamiltonian case and $x T / \delta_F$ in the Floquet case, the linear term has the same coefficient for both Hamiltonian and Floquet filters. 

\begin{figure}
	\centering
	\includegraphics[width=.4\textwidth]{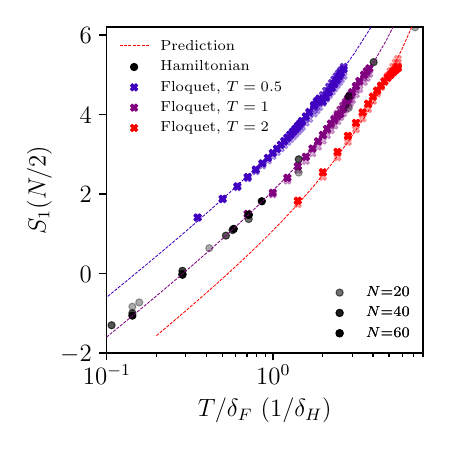}
	\caption{Scaling of half-chain von Neumann entanglement entropy of filtered states. In order to compare Hamiltonian and Floquet regimes, the $x$-axis shows $T / \delta_F$ in the Floquet case (crosses) and $ 1/  \delta_H$ in the Hamiltonian case (circles). Different shades correspond to  different total system sizes $N$. The dashed lines stand for entanglement entropies predicted by our theory in \cref{eq:floquet_vN_entropy}.
    We compute Floquet-filtered states with $T = 0.5, 1$, and $2$, represented by blue, purple, and red crosses respectively, using MPS simulations with bond dimension 500. 
    In the Hamiltonian case, we plot $S_1(\delta_H) - \log(\sqrt{N}) + C$, which according to \cref{eq:entropy_hamil} should coincide with the Floquet $T=1$ case. The fitted constant $C=0.58$. The data in the Hamiltonian regime were first reported in \cite{Banuls2020}. 
    }
	\label{fig:floquet_entropy}
\end{figure}

We performed MPS simulations to numerically check this equation, and the results are illustrated in \cref{fig:floquet_entropy}. 
The plot shows how, for Floquet filtered states with $T = 0.5, 1$, and $2$, and different total system sizes, the computed von Neumann entanglement entropies fit well with the theoretically derived form \cref{eq:floquet_vN_entropy}. For Hamiltonian-filtered states, we use the results first reported in \cite{Banuls2020}, which agree with \cref{eq:entropy_hamil}. In particular, the Hamiltonian case coincides with the Floquet case with $T = 1$ up to a $\varO(\log \sqrt{N})$ term produced by the entanglement entropy of the effective initial state $\ket{\Psi_n}$.
To make this visible in the figure, we first fit the entropies of the Hamiltonian case to $S_1(\delta_H) = \log(\sqrt{N}) + C$ and then plot the result of subtracting $\log(\sqrt{N}) + C$ from the entropy.

%% file: discussion.tex
\section{Summary and discussion}
\label{sec:summary_floquet_filter}
Energy filters can be expressed as a time series and applied on simple initial states, to obtain an approximation to microcanonical properties that improves as the filter width vanishes.
Here we provide a quantitative characterization of this convergence in terms of the filter width. 
To reach this result, we have addressed the thermality question for filtered states in the Floquet framework. 
In this setup, we consider stroboscopic applications of the evolution operator $U$, at time interval $T$, and the effect of the filter is reducing the variance, not in the energy, but in the  phase of $U$, which is proportional to the energies only for small enough $T$.

First, we showed the distinct scalings of R\'enyi-$\alpha$ entanglement entropies of Floquet-filtered states generated by local Hamiltonians:
\begin{itemize}
	\item when $\alpha > 1$, the R\'enyi entanglement entropy scales logarithmically in inverse filter width $1 / \delta$;
	\item when $\alpha = 1$, it is linear in $1 / \delta$;
	\item when $\alpha < 1$, it is quadratic in $1 / \delta$ with an exact cosine filter, and is linear in $1 / \delta$ with a truncated cosine filter.
\end{itemize}

The proof is based on the block-diagonal property that we found in the RDM of the filtered state. This leads to the equivalence of the Floquet filtered state expectation values and weighted time averages of given observables. The spectral filters provide a controllable suppression of off-diagonal matrix elements of observables in the ETH formalism, which helps to bound the difference between filtered and thermal states. We find a typical upper bound $\varO(\sqrt{\delta})$. This is in line with the  canonical universality property previously proposed in~\cite{Dymarsky2019}, but for filtered states it is no longer an independent hypothesis but can be derived from Floquet ETH. 

When $\delta$ is chosen to vanish in the thermodynamic limit, the filtered states $\ket{\psi_{\delta}}$ are a class of thermal pure quantum states (TPQS) that can typically approach the microcanonical ensembles with controllable error. The conventional methods to prepare TPQS~\citep{Sugiura2012,Sugiura2013,Iwaki2021} require a Haar random state as initial state, to eliminate the coherences between energy eigenstates by typicality arguments~\citep{Reimann2007}. In the filter construction, however, the initial state can instead be just a product state, which is simpler for both classical and quantum implementations.

Notably, filtered states in the Hamiltonian regime can be reduced to the Floquet case for initial product states and similar scalings hold. This reduction indicates that the Floquet filters can replace Hamiltonian filters in thermal state preparation, which provides a more efficient implementation as a quantum algorithm. The time step in the Floquet filter is $\varO(\sqrt{N})$ larger than that of a Hamiltonian filter, and hence the number of measurements are reduced by at least the same factor.
This behavior may be closely related to the phenomenon of prethermalization~\cite{Mori2018Prethermal}, which we leave for future work.

\begin{acknowledgments}
YY and JIC acknowledge the support from the German Federal Ministry of Education and Research (BMBF) through FermiQP (Grant No. 13N15890) and EQUAHUMO (Grant No. 13N16066) within the funding program quantum technologies - from basic research to market. This research is part of the Munich Quantum Valley (MQV), which is supported by the Bavarian state government with funds from the Hightech Agenda Bayern Plus. 
The work was also partially supported by the Deutsche Forschungsgemeinschaft (DFG, German Research Foundation) under Germany's Excellence Strategy -- EXC-2111 -- 390814868, and  SFB-TRR360 (project nr. 492547816). YY also acknowledges support from the U.S. Department of Energy, Office of Science, Accelerated Research in Quantum Computing Centers, Quantum Utility through Advanced Computational Quantum Algorithms, grant no. DE-SC0025572.
\end{acknowledgments}

%% file: appendix.tex
\begin{appendices}
\section{Normalized filter coefficients}
\label{appendix:filter_coef}
When applying the spectral filter \cref{eq:filter_cos_sum} to an initial state $\ket{\psi}$, the filtered state has to be normalized. The normalization factor for filter coefficients can be estimated as 
    \begin{equation}\begin{aligned}
        \mathcal{N}^2 = & \sum_{m,n=-xM}^{xM} c_m c_n \braket{\psi | U^{n - m} |\psi}\\
    \sim & \sum_{m,n = -xM}^{xM} c_m c_n e^{- N f((m-n) T)},
    \end{aligned}\end{equation}
where $f(t)$ is defined in \cref{eq:ft}.
\paragraph{Floquet regime.} In the Floquet regime, the terms for $m\neq n$ decay exponentially in system size, and hence 
\begin{equation}
    \mathcal{N}^2 \sim \sum_{m} c_m^2.
\end{equation}
Inserting the Sterling formula $n! \sim \sqrt{2\pi n} (n / e)^n  $, we obtain
\begin{equation}\begin{aligned}
    c_m & = \frac{1}{2^{2M^2}} \begin{pmatrix}
        2M^2 \\ M^2 - m
    \end{pmatrix}\\
    & = \frac{(2M^2)!}{2^{2M^2}(M^2 - m)! (M^2 + m)!} \\
    & \sim \frac{ h(u, M) }{\sqrt{\pi M^2}} ,
\end{aligned}\end{equation}
where $u = m / M$ and
\begin{equation}
    h(u, M) = \left[\frac{1}{\left(1-\frac{u}{M}\right)^{1-\frac{u}{M}} \left(1+\frac{u}{M}\right)^{1+\frac{u}{M}}}  \right]^{M^2}\cdot \frac{1}{\sqrt{1-\left(\frac{u}{M}\right)^2}}.
\end{equation}
In the limit $M \to \infty$, $h(u,M)$ pointwise converges to  $h_{\infty}(u) = \exp(-u^2)$. Thus
\begin{equation}\begin{aligned}
    \mathcal{N}^2 
    \sim &  \frac{1}{\pi M^2} \sum_{m=-xM}^{xM } h^2(u, M)\\
    \xrightarrow{M \to \infty} & \frac{1}{\pi M}  \int_{-x}^{x}\dInt u\,
     h_{\infty}^2(u)  \sim \frac{1}{\sqrt{2\pi} M}. \\
\end{aligned}\end{equation}
Hence the rescaled filter coefficient
\begin{equation}\begin{aligned}
    \tilde{c}_m^2 = & c_m^2 / \mathcal{N}^2 \approx \sqrt{2\pi}M \cdot \left(\frac{1}{2^{2M^2}} 
    \begin{pmatrix}
            2M^2 \\ M^2 - m
        \end{pmatrix} \right)^2 \\
    \sim & \frac{\sqrt{2\pi}}{\pi M} \exp\left(- 2m^2/M^2\right).
\end{aligned}\end{equation}
It is approximately a Gaussian function of width $M/2 = 1 / 2\delta_F$, and $\tilde{c}_0^2 \propto 1 / M = \delta_F$.

\paragraph{Hamiltonian regime.}
In the Hamiltonian regime,  we choose $T = p / \sigma_{\psi}\sqrt{N}$ for some constant $p$, where $\sigma_{\psi}^2 N$ is the energy variance of $\ket{\psi}$. 
When $t = \varO(1 / \sqrt{N})$, i.e., when the Loschmidt echo is not yet exponentially small in system size, we have $f(t) \sim \sigma_{\psi}^2t^2 / 2$~\cite{Hartmann2004} for $f$ defined in \eqref{eq:ft}, we can further approximate the normalization factor as 
\begin{align}
    \mathcal{N}^2 \sim  \sum_{m,n = -xM}^{xM} c_m c_n e^{-\sigma_{\psi}^2 N (m-n)^2 T^2 / 2}.
\end{align}

Let $v = n / M$, and we get 

\begin{equation}\begin{aligned}
    \mathcal{N}^2 
    \sim &  \frac{1}{\pi M^2} \sum_{m,n=-xM}^{xM} e^{-\frac{p^2M^2(u-v)^2}{2}} h(u, M) h(v, M)\\
    \xrightarrow{M \to \infty} & \frac{1}{\pi}  \int_{-x}^{x}\dInt u\,\dInt v\, \lim_{M\to\infty} e^{-\frac{p^2M^2(u-v)^2}{2}} h_{\infty}(u) h_{\infty}(v)\\
    = & \frac{1}{\pi}  \int_{-x}^{x}\dInt u\,\dInt v\, \frac{\sqrt{2\pi}}{pM}\delta(u-v) h_{\infty}(u) h_{\infty}(v)\\
    = & \frac{\sqrt{2\pi}}{\pi pM} \int_{-x}^{x} \dInt u\, h^2_{\infty}(u) \sim \frac{1}{pM}.
\end{aligned}\end{equation}
Hence the rescaled filter coefficient
\begin{equation}\begin{aligned}
    \tilde{c}_m = & c_m / \mathcal{N} \approx \sqrt{pM} \cdot \frac{1}{2^{2M^2}} 
    \begin{pmatrix}
            2M^2 \\ M^2 - m
        \end{pmatrix}  \\
    \sim & \sqrt{\frac{p}{\pi M} }\exp\left[- \left(\frac{m}{M}\right)^2\right] = \varO\left( 1 / \sqrt{M}\right).
\end{aligned}\end{equation}

\section{Derivation of \cref{eq:ortho_diag}}
\label{appendix:proof_ortho}
In this section, we show that, for $m\neq n$,
\begin{equation}   \tr\left( \rho_{A,mm} \rho_{A,nn} \right) \le \alpha_{\delta_F} 2^{- \beta_{\delta_F} N_A},
\end{equation}
where the constants $\alpha_{\delta_F}$ and $\beta_{\delta_F}$  depend only on the model and the filter width $\delta_F$.

This result is established under the assumption, introduced in \cref{eq:gamma_nT}, that the Loschmidt echoes of all product states (which need not be translationally invariant) are uniformly bounded by a quantity exponentially small in the system size $N$ at all times $nT$, for any integer $n\neq 0$. Namely, there exists $\gamma_n > 0$ such that for any product state $\ket{\psi}$,
\begin{equation}
\left|\braket{\psi | e^{-inHT} | \psi}\right|^2 \le 2^{-N \gamma_n}.
\label{eq:assump_loschmidt}
\end{equation}

Let us rewrite the Hamiltonian as 
\begin{equation}
    H = H_0 + H_{\partial},
\end{equation}
where $H_0 = H_A + H_B$, $H_{A(B)}$ denotes the sum of Hamiltonian terms in $H$ acting solely on subsystem $A(B)$ and $H_{\partial}$ represents the boundary term. 
The evolution operator $U^n$ can be written as
\begin{equation}
    U^n = e^{-iHnT} = e^{-iH_0 nT} V(nT),
\end{equation}
where
\begin{equation}
    V(t):= \mathcal{T} \exp\left(-i \int_0^t e^{iH_0 s} H_{\partial} e^{-i H_0 s} \dInt s \right)
\end{equation}
is the evolution in the interaction picture with respect to $H_0$.

\begin{figure}
    \centering
    \includegraphics[width=0.9\linewidth]{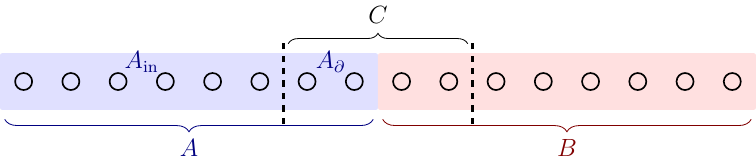}
    \caption{Factorization of the system.}
    \label{fig:factorization}
\end{figure}

According to the Lieb-Robinson bound, $V(t)$ is quasi-local, in the sense that it can be approximated by a local unitary
\begin{equation}
    V'(t):= \mathcal{T} \exp\left(-i \int_0^t e^{i(H_0)_{C} s} H_{\partial} e^{-i (H_0)_{C} s} \dInt s \right).
\end{equation}
where $C$ is an area centered at the boundary (see \cref{fig:factorization}), and $(H_0)_C$ is the restriction of $H_0$ to this region. It holds that~\cite{Osborne2006Lieb}
\begin{equation}
    \norm{V(t) - V'(t)} \le \omega e^{\kappa |t|}2^{-\mu N_C},
\end{equation}
where $\omega$, $\kappa$ and $\mu$ are constants independent of the system size. Let us choose $N_C = cN_A \propto N_A$ for some constant $c$, so that $\norm{V(t) - V'(t)} = e^{-\Omega(N_A)}$ for $t = o(N_A)$.

Since $H_A$ and $H_B$ act purely on subsystems $A$ and $B$, respectively, we have
\begin{equation}
    \begin{aligned}
        & \rho_{A,nn} = \tr_B \left( U^n \ket{\psi}\bra{\psi} U^{\dagger n} \right) \\
        = & e^{-iH_A nT} \tr_B \left( V(nT) \ket{\psi} \bra{\psi} V^{\dagger}(nT)\right)  e^{iH_A nT}.
    \end{aligned}
\end{equation}
Let us define the approximate reduced density matrix
\begin{equation}
    \begin{aligned}
        \rho'_{A,nn} :&= e^{-iH_A nT} \tr_B \left( V'(nT) \ket{\psi} \bra{\psi} V'^{\dagger}(nT)\right)  e^{iH_A nT}\\
        & = e^{-iH_A nT} \left( \ket{\psi_{A_{\rm in}}}\bra{\psi_{A_{\rm in}}} \otimes \sigma_{n} \right) e^{iH_A nT},
    \end{aligned}
    \label{eq:rhonn}
\end{equation}
where $A_{\rm in}:=A \backslash C$, and 
\begin{equation}
    \sigma_n:= \tr_{B\cap C}\left( V'(nT) \ket{  \psi_{C}}\bra{\psi_C}V'^{\dagger}(nT) \right)
    \label{eq:rhonnprime}
\end{equation}
is a reduced density matrix supported on $A_{\partial } := A \cap C$. According to \cref{eq:rhonn} and the first line of \cref{eq:rhonnprime}, the trace distance
\begin{equation}
    \begin{aligned}
        \norm{\rho_{A,nn} - \rho'_{A,nn}}_1
        \le & 2 \norm{V(nT) - V'(nT)}\\
        \le & 2\omega e^{n \kappa T}2^{-\mu c N_A},
    \end{aligned}
\end{equation}
and thus we can bound
\begin{widetext}
\begin{equation}
    \begin{aligned}
         \left| \tr\left( \rho_{A,mm} \rho_{A,nn} \right) - \right. & \left. \tr\left( \rho'_{A,mm} \rho'_{A,nn} \right) \right| \\
        \le & \left| \tr\left[ \rho_{A,mm}   \left( \rho_{A,nn} -  \rho'_{A,nn} \right) \right] \right| + \left| \tr\left[  \left( \rho_{A,mm} -  \rho'_{A,mm} \right)  \rho'_{A,nn}\right]  \right|\\
        \le & 2\omega( e^{n \kappa T} + e^{m \kappa T})2^{-\mu c N_A}.
    \end{aligned}
    \label{eq:trace_approx}
\end{equation}

We then compute 
\begin{equation}
    \begin{aligned}
        \tr\left( \rho'_{A,mm} \rho'_{A,nn} \right) & = \tr\left( e^{-iH_A (n-m)T} \left( \ket{\psi_{A_{\rm in}}}\bra{\psi_{A_{\rm in}}} \otimes \sigma_{n} \right) e^{iH_A (n-m)T} \left( \ket{\psi_{A_{\rm in}}}\bra{\psi_{A_{\rm in}}} \otimes \sigma_{m} \right) \right)\\
        & \le \tr\left( e^{-iH_A (n-m)T} \left( \ket{\psi_{A_{\rm in}}}\bra{\psi_{A_{\rm in}}} \otimes \Id_{{A_{\partial }}} \right) e^{iH_A (n-m)T} \left( \ket{\psi_{A_{\rm in}}}\bra{\psi_{A_{\rm in}}} \otimes \Id_{{A_{\partial }}} \right) \right)\\
        & = \sum_{1 \le i,j \le d_{{A_{\partial }}}} \left| \underbrace{\bra{\psi_{A_{\rm in}} \otimes i_{A_{\partial }} } }_{\bra{\Psi_i}} e^{iH_A (n-m)T} \underbrace{\ket{\psi_{A_{\rm in}} \otimes j_{A_{\partial }} }   }_{\ket{\Psi_j}}\right|^2.
    \end{aligned}
\end{equation}
\end{widetext}
where $d_{{A_{\partial }}} = 2^{N_C/2}$ is the Hilbert space dimension of the subsystem ${A_{\partial }}$, and $\left\{\ket{i_{A_{\partial }}}\right\}_{1\le i \le d_{A_{\partial }}}$ is a product basis set. 
For single qubit states $\ket{a}, \ket{b}$, we can write $\ket{a}\bra{b}$ into a sum of projectors as
\begin{equation}
    \begin{aligned}
    \ket{a}\bra{b} & = \frac{1}{4}\sum_{k=0}^3 i^k \left(\ket{a}+i^k \ket{b}\right)\left(\bra{a}+(-i)^k \bra{b}\right)\\
    & = \frac{1}{4}\sum_{k=0}^3 c_k^{a,b} P_{k}^{a,b}.
    \end{aligned}
\end{equation}
The normalization coefficient satisfies $|c_k^{a,b}| \le 2$.
Applying the decomposition to all sites in $A_{\partial }$, we get
\begin{equation}
    \begin{aligned}
        \ket{j_{A_{\partial }}}\bra{i_{A_{\partial }}} = \frac{1}{4^{N_{A_{\partial }}}} \sum_{\bm{k}\in \left\{0,1,2,3\right\}^{\otimes N^{A_{\partial }}}}  c_{\bm{k}}^{i_{A_{\partial }},j_{A_{\partial }}} P_{\bm{k}}^{i_{A_{\partial }},j_{A_{\partial }}},
    \end{aligned}
\end{equation}
where $P_{i_{A_{\partial }},j_{A_{\partial }},\bm{k}}$ are projectors of product states in subsystem $A_{\partial }$. Therefore
\begin{equation}
\begin{aligned}
    & \left| \bra{\Psi_i} e^{-iH_A (m-n)T} \ket{\Psi_j } \right| = \left| \Tr \left(  e^{-iH_A (m-n)T} \ket{\Psi_j } \bra{\Psi_i} \right)\right|\\
    \le & \frac{2^{N_{A_{\partial }}}}{4^{N_{A_{\partial }}}} \sum_{\bm{k}}\left| \Tr \left( e^{-iH_A (m-n)T} \ket{\psi_{A_{\rm in}}}\bra{\psi_{A_{\rm in}}}\otimes P_{\bm{k}}^{i_{A_{\partial }},j_{A_{\partial }}} \right)\right|\\
    \le & 2^{N_{A_{\partial }}} 2^{-N_A \gamma_{m-n}}.
\end{aligned}
\end{equation}
The last step follows from assumption \cref{eq:assump_loschmidt}. Therefore
\begin{equation}
    \begin{aligned}
    & \tr\left( \rho'_{A,mm} \rho'_{A,nn} \right) \le d_{{A_{\partial }}}^3 2^{-N_A \gamma_{m-n}} = 2^{-N_A (\gamma_{m-n} - 3c/2 )}.
    \end{aligned}
\end{equation}

Together with \cref{eq:trace_approx}, we can choose the constant $c=N_C / N_A = \gamma_{m-n} / ( \mu + 3 / 2)$ such that
\begin{equation}
    \tr\left( \rho_{A,mm} \rho_{A,nn} \right) \le \left(1 + 2\omega (e^{n \kappa T} + e^{m\kappa T})\right)2^{-\frac{\gamma_{m-n}} {1 + 3 / 2 \mu} N_A}.
\end{equation}
Let $\alpha_{\delta_F}:=1 + 4\omega e^{n\kappa xM}$, and $\beta_{\delta_F} : = \frac{\gamma_{\min}}{1 + 3 / 2 \mu }$. 
Here $\gamma_{\min} = \min_{|n| \le xM} \gamma_{n}$ is the smallest value of $\gamma_n$ that can occur with the given filter width. Finally, we obtain
\begin{equation}
    \tr\left( \rho_{A,mm} \rho_{A,nn} \right) \le \alpha_{\delta_F} 2^{- \beta_{\delta_F} N_A}.
\end{equation}

\section{Eigenvalues of $\rho_{A,\delta_F}$}
\label{appendix:eigvals}
In this appendix section, we demonstrate why the eigenvalues of $\rho_{A,\delta_F}$ can be given by the collections of nonzero eigenvalues of $\rho_{A, mm}$ rescaled by $\tilde{c}_m^2$, as claimed in the end of \cref{sec:rdm_floquet}, and why they can be used to compute the R\'enyi entanglement entropies of $\rho_{A,\delta_F}$.

Let us denote the eigenvalues of $\rho_{A,\delta_F}$ by $s_k$, those of $\sum_{m}\tilde{c}_m^2 \rho_{A,mm}$ by $s'_k$. Let $s''_k$ be the collection of nonzero eigenvalues of $\rho_{A,mm}$ rescaled by $\tilde{c}_m^2$. We assume that $s_k$, $s'_k$, and $s''_k$ are all sorted in a descending order. We first show that 
\begin{equation}
    |s_k - s_k''| = e^{-\Omega(N_A)},
\end{equation}
when both conditions in \cref{sec:rdm_floquet} are satisfied. The proof consists of two steps:
\begin{enumerate}
    \item $|s_k - s_k'| = e^{-\Omega(N_A)}$.
    It directly follows from the Hoffman-Wielandt inequality:
    \begin{equation}
    \begin{aligned}
        \sum_{k} \left(s_k-s_{k}'\right)^2
        \le & \left\| \rho_{A,\delta_F} - \sum_{m}\tilde{c}_m^2 \rho_{A,mm} \right\|_F^2 = e^{-\Omega(N_B)}.
    \end{aligned}
    \end{equation}
    Since $N_B \gg N_A$, if follows that $|s_k - s_k'| = e^{-\Omega(N_A)}$.
    \item $|s_k'-s_k''| = e^{-\Omega(N_A)}$. 
    Let us define a block matrix
    \begin{equation}
        T = \begin{pmatrix}
            T_{-xM}\\T_{-xM+1}\\\vdots\\T_{xM}
        \end{pmatrix}
    \end{equation}
    with $T_{m} = \tilde{c}_m\sqrt{\rho_{A,mm}}$. Then we have
    \begin{equation}
        \begin{aligned}
            T^{\dagger} T &= \sum_{m} \tilde{c}_m^2 \rho_{A,mm},\\
            (TT^{\dagger})_{mn} & = \tilde{c}_m\tilde{c}_n \sqrt{\rho_{A,mm} \rho_{A,nn}}.
        \end{aligned}
    \end{equation}
    $TT^{\dagger}$ share the same nonzero eigenvalues with $T^{\dagger}T$, which are $s'_k$. Let $R$ be a block-diagonal matrix that contains the diagonal blocks of $TT^{\dagger}$
    \begin{equation}
        R = {\rm diag} \left\{ \tilde{c}_m^2 \rho_{A,mm} \right\}.
    \end{equation}
    The nonzero eigenvalues of $R$ are exactly $s''_k$. According to the Weyl's inequality,
    \begin{equation}
        \begin{aligned}
            |s'_k-s''_k| \le &\left\| TT^{\dagger} - R \right\|
            \le \left\| TT^{\dagger} - R \right\|_F \\
            = & \sqrt{\sum_{m\neq n} \tilde{c}_m^2 \tilde{c}_n^2 \tr \left( \rho_{A,mm} \rho_{A,nn} \right) } = e^{-\Omega(N_A)}.
        \end{aligned}
    \end{equation}
    
\end{enumerate}

The rest is to bound the error in estimating R\'enyi-$\alpha$ entropy of $\rho_{A, \delta_F}$ with the approximated eigenvalues $\{s_k''\}$.
\begin{itemize}
    \item When $\alpha > 1$,
\begin{equation}
    |s_k^{\alpha} - s_{k}''^{\alpha}|\le \left(\sup_{x\in [0,1]} \frac{\dInt x^{\alpha}}{\dInt x} \right) |s_k-s_k''|=\alpha |s_k-s_k''|.
\end{equation}
Since the R\'enyi entanglement entropy $S_{\alpha}(\rho_{A,mm})$ grows at most linearly in $m$~\cite{Toniolo2025}, ${\rm rank}(\rho_{A,mm}) = e^{\varO(m)}$ and hence the rank $\mathcal{D}$ of $\rho_{A,\delta_F}$ is bounded by $\mathcal{D} = e^{\varO(1 / \delta_F)}$. When $1/\delta_F= o (N_A)$, 
\begin{equation}
    \left|\sum_{k}s_k^{\alpha} - \sum_{k}s_k''^{\alpha}\right| \le \mathcal{D} \alpha \max_{k} |s_k-s_k''| = e^{-\Omega(N_A)}.
\end{equation}
According to the calculations in \cref{sec:mechanism_filter},
\begin{equation}
    S_{\alpha>1}\left(\sum_{m}\tilde{c}_m^2 \rho_{A,mm} \right) = o(N_A),
\end{equation}
for $1/ \delta_F = o(N_A)$, and $\sum_{k} s_k''^{\alpha} = e^{-o(N_A)} $. Therefore
\begin{equation}
    \begin{aligned}
        & S_{\alpha}(\rho_{A,\delta_F}) - S_{\alpha}\left( \{ s_k'' \} \right)\\
        = & \frac{1}{1-\alpha}\log \frac{\sum_k s_k^{\alpha}}{\sum_k s_k''^{\alpha}} \\
        = &  \frac{1}{1-\alpha}\log \left( 1 + \frac{\sum_k s_k^{\alpha} - \sum_k s_k''^{\alpha}}{\sum_k s_k''^{\alpha}} \right) = e^{-\Omega(N_A)}.
    \end{aligned}
\end{equation}

\item In the case $0<\alpha<1$, due to the concavity of $x^{\alpha}$, we have 
\begin{equation}
    \left(x+(y-x)\right)^{\alpha} \le x^{\alpha} + (y-x)^{\alpha},
\end{equation}
or $y^{\alpha} - x^{\alpha} \le (y-x)^\alpha$
for $0<x<y$, and thus 
\begin{equation}
    |s_k^{\alpha} - s_k''^{\alpha}| \le | s_k - s_k'' |^{\alpha}.
\end{equation}
Then same analysis then follows as the $\alpha > 1$ case.

\item 
Finally, for $\alpha = 1$, according to the Fannes–Audenaert inequality~\cite{Audenaert2007},
\begin{equation}
    \begin{aligned}
        & \left|S_1\left(\rho_{A,\delta_F}\right) - S_1\left( \{ s_k'' \} \right)\right|\\
        \le & \delta \log (\mathcal{D}-1) +S_1(\{\delta, 1 - \delta\}),
    \end{aligned}
\end{equation}
where $\delta = \sum_k |s_k-s_k''| / 2 = e^{- \Omega(N_A)}$. Therefore $\left|S_1\left(\rho_{A,\delta_F}\right) - S_1\left( \{ s_k'' \}  \right)\right| = e^{-\Omega(N_A)}$ also holds.

\end{itemize}

\section{Bounding the higher order moments of $Q(\delta_F)$}
\label{appendix:q_delta}
In \cref{sec:filter_thermal}, we want to compute $\tr\left[ Q(\delta_F)^{2l} \right]$. As a warm up, in $l=1$ case, we have
\begin{equation}\begin{aligned}
    \overline{\tr\left[ Q(\delta_F)^2 \right]} & = \sum_{k,k'} \overline{|R_{k,k'}|^2}  e^{-\frac{(\phi_k - \phi_{k'})^2}{4\delta_F^2}}/ \mathcal{D}\\
    & \approx \frac{\mathcal{D}}{2\pi}  \int_{-\pi}^{\pi} \dInt \varphi\, e^{-\frac{\varphi^2}{4\delta_F^2}} \approx  \frac{\mathcal{D}}{\sqrt{\pi}} \delta_F.
\end{aligned}\end{equation}
Here we assume an even distribution of eigenvalues in the periodic Floquet spectrum $[-\pi, \pi)$ and $\delta_F \ll \pi$. The summation over $k$ produces a factor $\mathcal{D}$, and summation over $k'$ is approximated by $\sum_{k'} \to \frac{\mathcal{D}}{2\pi}  \int_{-\pi}^{\pi} \dInt \varphi$, where $\varphi$ represents the phase difference $\phi_{k} - \phi_{k'}$. 

The higher order moments contain contribution of off-diagonal elements $R_{k,k'}$ in ETH ansatz. To perform the calculation, we additionally assume that ~\cite{Foini2019,Foini2019a}
\begin{eqnarray}
    \overline{ R_{k_1,k_2} R_{k_2,k_3}\cdots R_{k_n,k_1} } = F_n / \mathcal{D}^{\frac{n}{2}-1}.
\end{eqnarray}
For example, the expectation value of the fourth order correlation function of $R_{k,k'}$ can be written as
\begin{equation}\begin{aligned}
    & \overline{R_{k_1,k_2}R_{k_2,k_3}R_{k_3,k_4}R_{k_4,k_1}} \\
    = &  \delta_{k_1,k_3} \overline{R_{k_1,k_2} R_{k_2,k_1}} \cdot \overline{R_{k_1,k_4} R_{k_4,k_1}}\\
    &+ \delta_{k_2,k_4} \overline{R_{k_1,k_4} R_{k_4,k_1}} \cdot \overline{R_{k_3,k_4} R_{k_4,k_3}}\\
    & + \eta_{k_1,k_2,k_3,k_4} \overline{R_{k_1,k_2}R_{k_2,k_3}R_{k_3,k_4}R_{k_4,k_1}}\\
    =& \delta_{k_1,k_3}  + \delta_{k_2,k_4} + \eta_{\bm{k}} F_4 / \mathcal{D},
\end{aligned}\end{equation}
where $\eta_{\bm{k}}$ takes value 1 when all the indices of $\bm{k}$ are different, and otherwise 0. Now we can compute the fourth order moment:
\begin{equation}\begin{aligned}
     & \overline{\tr\left[ Q(\delta_F)^4 \right]} \\
     = & \frac{1}{\mathcal{D}^2} \sum_{\bm{k}} \overline{R_{k_1,k_2}R_{k_2,k_3}R_{k_3,k_4}R_{k_4,k_1}} e^{-\frac{\phi_{12}^2 +\phi_{23}^2+ \phi_{34}^2 + \phi_{41}^2}{8\delta_F^2}}\\
    \sim & \frac{1}{\mathcal{D}^2} \sum_{\bm{k}} \left[ \delta_{k_1,k_3} + \delta_{k_2, k_4} + \eta_{\bm{k}} {F_4} / {\mathcal{D}} \right] e^{-\frac{ \phi_{12}^2 +  \phi_{23}^2+ \phi_{34}^2 + \phi_{41}^2}{8\delta_F^2}},
\end{aligned}\end{equation}
where $\phi_{ij} = \phi_{k_i} - \phi_{k_j}$. The $\delta_{k_1, k_3}$ term can be computed as
\begin{equation}
    \frac{1}{\mathcal{D}} \left( \frac{\mathcal{D}}{2\pi} \right)^2 \int \dInt \phi_{12}  \dInt \phi_{14}\, e^{-\frac{\phi_{12}^2 + \phi_{14}^2}{4\delta_F^2}}  \approx \frac{\mathcal{D}}{\pi} \delta_F^2.
\end{equation}
The $\delta_{k_2,k_4}$ term is analogous. The term with higher order correlations of $R_{k,k'}$ gives
\begin{equation}
\begin{aligned}
    &\frac{F_4}{\mathcal{D}^2} \left(\frac{\mathcal{D}}{2\pi} \right)^3 \int \dInt \phi_{12} \dInt \phi_{23}  \dInt \phi_{34}\, e^{-\frac{ \phi_{12}^2 + \phi_{23}^2 + \phi_{34}^2 + (\phi_{12}+\phi_{23} + \phi_{34})^2 }{8\delta_F^2}} \\
    & \approx  \frac{\sqrt{2}F_4}{\pi^{3/2}}\mathcal{D}\delta_F^3 = \varO(\mathcal{D} \delta_F^3),
\end{aligned}
\end{equation}
which only contributes to higher order corrections in $\delta_F$, compared with the two point correlation terms. Therefore $\tr\left[ Q(\delta_F)^{4}\right] = \varO(\mathcal{D}\delta_F^{2})$. 

The computation with higher order ($2l$) moments are analogous. When the indices of $\overline{ R_{k_1,k_2} R_{k_2,k_3}\cdots R_{k_{2l},k_1} }$ are taken such that they form $l$ conjugated pairs of the form $\overline{R_{m,n}R_{n,m}}$, the corresponding integrals will contribute a  $\varO(\mathcal{D}\delta_F^l)$ term. When the indices form longer loops, they will end up with higher order corrections in $\delta_F$.

\section{Trace distance between Hamiltonian-filtered state and its Floquet-filtered state approximation}
\label{appendix:diff_states}

In this appendix section, we bound the trace distance between the Hamiltonian-filtered state $\ket{\psi_{\delta_H}}$ and its Floquet-filtered state approximation 
\begin{eqnarray}
	\ket{\psi_{\delta_H}} \approx  \ket{\tilde{\psi}_{\delta_H}} = \sum_{k = - \floor{xM / n - 1 / 2}}^{\floor{xM / n - 1 / 2}} \tilde{c}_{kn} U^{k n}  \ket{\Psi_n},
\end{eqnarray}
as stated in \cref{sec:floquet_to_hamil}.

As already shown in \cref{appendix:filter_coef}, the rescaled filter coefficient $\tilde{c}_m  = \varO( 1 / \sqrt{M})$.
Therefore for any $\Delta = \varO(\sqrt{N})$, according to the definition in \eqref{eq:def_cm},
\begin{equation}\begin{aligned}
   & |\tilde{c}_m - \tilde{c}_{m+\Delta}| = \tilde{c}_m \left| 1 - \prod_{k=1}^{\Delta} \frac{ M^2 - m - \Delta + k }{ M^2 + m + k } \right|\\
= & \tilde{c}_m \left| 1 - \prod_{k=1}^{\Delta} \left(1 - \frac{  2 m - \Delta}{ M^2 + m + k } \right)  \right|\\
 = & \tilde{c}_m \cdot \varO\left(  \frac{(2m - \Delta)\cdot\Delta}{M^2}  \right) = \varO\left( {\Delta} / {M^{3 / 2}} \right).
\end{aligned}\end{equation}
The Euclidean distance between states $\ket{\psi_{\delta_H}}$ and $\ket{\tilde{\psi}_{\delta_H}}$, or the squared norm of $\ket{q} = \ket{\psi_{\delta_H}} - \ket{\tilde{\psi}_{\delta_H}}$, is then bounded by 
\begin{widetext}
\begin{equation}\begin{aligned}
    \norm{ \ket{q} }^2 = & \sum_{k,k'}\sum_{m,m'} (\tilde{c}_{kn+m}-\tilde{c}_{kn}) (\tilde{c}_{k'n+m'}-\tilde{c}_{k'n}) \braket{\psi | U^{(k-k')n+(m-m')} | \psi}\\
    \le & \sum_{k,k'}\sum_{m,m'} |(\tilde{c}_{kn+m}-\tilde{c}_{kn}) (\tilde{c}_{k'n+m'}-\tilde{c}_{k'n}) | \cdot |\braket{\psi | U^{(k-k')n+(m-m')} | \psi} | \\
    \le & \sum_{|kn+m-(k'n+m')| \le K\log N } | (\tilde{c}_{kn+m}-\tilde{c}_{kn}) (\tilde{c}_{k'n+m'}-\tilde{c}_{k'n}) |  + \varO(M^2) \cdot e^{-\varO(N^{2\epsilon})}\\
    = & \varO( M \log N \cdot (n / M^{3/2})^2 ) = \varO( \delta_H^2 \log N ) = \tilde{\varO}(\delta_H^2),
\end{aligned}\end{equation}
\end{widetext}
where $K > 0$ is a positive constant. 
In the second inequality, the summation is divided into two parts. The Loschmidt echo is upper bounded by 1, which is taken when $|kn+m-(k'n+m')| \le K \log N$. When $|kn+m-(k'n+m')| > K \log N$, we can again take the advantage of the fact that the Loschmidt echo of a product state $\ket{\psi}$ decays as a Gaussian in time around $t = 0$~\cite{Hartmann2004}: $|\braket{\psi | e^{-iHt} | \psi} | \sim e^{ -\sigma_{\psi}^2 N t^2 / 2 }$, where $\sigma_\psi^2 N$ is the energy variance of $\ket{\psi}$. Insert $t = |kn+m-(k'n+m')| T$ with $T = \varO(1 / \sqrt{N})$, and we obtain:
\begin{align}
    |\braket{\psi | U^{(k-k')n+(m-m')} | \psi} | \lesssim  e^{-\varO(\log^2 N)}.
\end{align} 
These terms thus only have sub-exponentially small contributions when $M = {\rm poly} (N)$~\footnote{Here we again assume that there is no significant revival of the Loschmidt echo in later times.}.

The normalization factor for $\ket{\tilde{\psi}_{\delta_H}}$ is $\tilde{\mathcal{N}}^2 =  \braket{\tilde{\psi}_{\delta_H} | \tilde{\psi}_{\delta_H}} = (\bra{\psi_{\delta_H}} - \bra{q})(\ket{ \psi_{\delta_H} } - \ket{q}) = 1 + \varO(\norm{\ket{q}})$, and thus we can estimate the trace distance $d$ between $\ket{\psi_{\delta_H}}$ and $\ket{\tilde{\psi}_{\delta_H}} / \tilde{\mathcal{N}}$:
\begin{equation}\begin{aligned}
    d & = \sqrt{1 - |\braket{ \psi_{\delta_H} | \tilde{\psi}_{\delta_H} }|^2  / \tilde{\mathcal{N}}^2 }\\
    & = \varO( \sqrt{\norm{ \ket{q} }}) = \tilde{\varO}(\sqrt{\delta_H} ).
\end{aligned}\end{equation}

\section{Two exactly solvable models}
\label{appendix:solvable_eg}
In this appendix section, we present two different  specific exactly solvable models, a Floquet toy model and in a more extreme setup, the global random matrix model to illustrate our results. We will compute their reduced density matrices, which validates the arguments for scalings of R\'enyi-entanglement entropies and thermality.

\subsection{A Floquet toy model}
\label{appendix:floquet_toy}
\begin{figure}
	\centering
	\includegraphics[width = .4\textwidth]{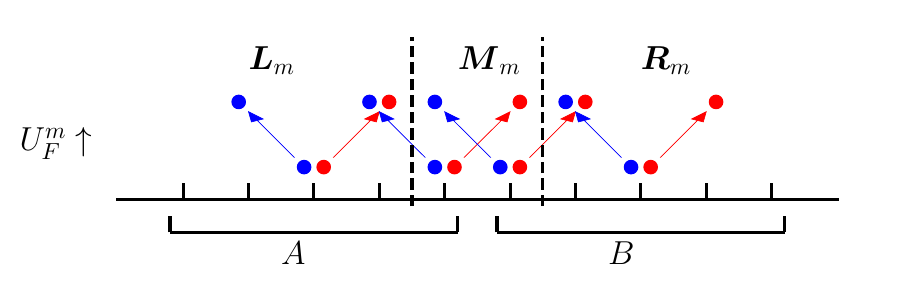}
	\caption{Evolution of the Floquet toy model. The blue points represent $l$ excitations and the red ones stand for $r$ excitations. After applying Floquet unitary operator, $l$ excitations move to the left and the $r$ excitations move to the right. The system is divided into three regions $\bm{L}_m$, $\bm{M}_m$ and $\bm{R}_m$ such that excitations initially at $\bm{L}_m$ and $\bm{R}_m$ will remain in subsystems $A$ and $B$, respectively, after applying $U_F^m$. While initial $l$ excitations in $\bm{M}_m$ will move to subsystem $A$, and $r$ excitations will move to subsystem $B$. The length of $\bm{M}_m$ is $2m$.}
	\label{fig:toy_model_move}
\end{figure}

\paragraph{The model.}
Let us think about a simple Floquet system that can be used to mimic the dynamics of freely propagating quasi-particles. Consider a chain of length $2N$ with sites labeling from $-N /2 +1 $ to $3N / 2$, where on each site $i$ there can be two degrees of freedom, the left- and right- moving ``excitations'' $l_i$ and $r_i$. The local complete basis set is therefore $\set{ \ket{0}, l_i^{\dagger}\ket{0}, r_i^{\dagger}\ket{0}, l_i^{\dagger}r_i^{\dagger}\ket{0} }$, where $\ket{0}$ is the vacuum state. The system is evolved under Floquet operator $U_F$, such that
\begin{equation}
	U_F l_i^{\dagger} U_F^{\dagger} = l_{i-1}^{\dagger}, \quad U_F r_i^{\dagger} U_F^{\dagger} = r_{i+1}^{\dagger}.
\end{equation}
That is, the $l_i$ excitations will move one site to the left, while $r_i$ ones will move one site to the right. \cref{fig:toy_model_move} provides a graphical illustration of the model.
The initial state is chosen to be a product of entangled $l$, $r$ pairs on sites $1, \cdots, N$:
\begin{equation}
	\ket{\psi_0} \propto \prod_{i=1}^{N} (\id + l_i^{\dagger} r_i^{\dagger}) \ket{0}^{\otimes 2N}.
	\label{eq:ini_toymodel}
\end{equation}
Below we will investigate how the half chain entanglement entropy grow as the function of Floquet filtered width $\delta_F$. Note that we focus on the regime where $1 \ll M = {1} / {\delta_F} \ll N$.

Let us use a more compact notation
\begin{equation}
	\ket{l_{\bm{s}} r_{\bm{s}'}} = \prod_{i\in \bm{s}} l_i^{\dagger} \prod_{j\in \bm{s}'} r_{j}^{\dagger} \ket{0}^{\otimes 2N},
\end{equation}
Then the initial state in \cref{eq:ini_toymodel} can be rewritten as 
\begin{equation}
	\ket{\psi_0} = \sum_{\bm{s} \subset \set{1,\cdots,N}} \ket{l_{\bm{s}} r_{\bm{s}}}.
\end{equation}
After $m$ time steps, the state $\ket{l_{\bm{s}} r_{\bm{s}}}$ will evolve to be $\ket{l_{\bm{s}-m}, r_{\bm{s}+m}}$, where $\bm{s}\pm m = \set{ i\pm m | \forall i \in \bm{s} }$. When considering the bipartition of the system into left and right halves, in case $m \ge 0$,
\begin{itemize}
	\item if the initial position of a pair of excitations $i \le N/2 - m$, these excitations will remain in the left half of the system. We call this region of the system $\bm{L}_m = \set{-L/2+1,\cdots,N/2-m}$;
	\item if $N/2 - m + 1\le i \le N / 2 + m$, the left excitation will move to the left half, while the right excitation will move to the right half. We call this region of the system $\bm{M}_m = \set{N/2-m+1,\cdots N/2+m}$;
	\item if $i > N / 2 + m$, the pair of excitation will remain in the right half of the system. We call this region of the system $\bm{R}_m = \set{N/2+m+1,\cdots,3N/2}$.
\end{itemize}
With such division of the system (see \cref{fig:toy_model_move}), the evolution of the state $\ket{\bm{l}_s\bm{r}_s}$ can be written as
\begin{equation}\begin{aligned}
    & U^m \ket{l_{\bm{s}} r_{\bm{ s }}} \\
    = & \ket{l_{(\bm{s}\cap{\bm{L}_m})-m}r_{(\bm{s}\cap{\bm{L}_m})+m}; l_{(\bm{s}\cap{\bm{M}_m})-m}}_A\\
    & \otimes \ket{r_{(\bm{s}\cap{\bm{M}_m})+m}; l_{(\bm{s}\cap{\bm{R}_m})-m}r_{(\bm{s}\cap{\bm{R}_m})+m}}_B
\end{aligned}\end{equation}
where $\bm{s}\cap{\bm{L}_m}$, $\bm{s}\cap{\bm{M}_m}$, and $\bm{s}\cap{\bm{R}_m}$ represent the components of $\bm{s}$ lying in regions $\bm{L}_m$, $\bm{M}_m$, and $\bm{R}_m$, respectively. If $m < 0$, the $r$ excitations instead move to the left and $l$ excitations to the right. Hence the evolved configuration is 
\begin{equation}\begin{aligned}
    & U^m \ket{l_{\bm{s}} r_{\bm{ s }}} \\
    = & \ket{l_{(\bm{s}\cap{\bm{L}_{|m|}})+|m|}r_{(\bm{s}\cap{\bm{L}_{|m|}})-|m|}; r_{(\bm{s}\cap{\bm{M}_{|m|}})-|m|}}_A\\
    & \otimes \ket{l_{(\bm{s}\cap{\bm{M}_{|m|}})+|m|}; l_{(\bm{s}\cap{\bm{R}_{|m|}})+|m|}r_{(\bm{s}\cap{\bm{R}_{|m|}})-|m|}}_B
\end{aligned}\end{equation}

\paragraph{The reduced density matrix of the filtered state.}
Using the above notations, we can compute the reduced density matrix (RDM) of the filtered state for the left half of the whole system $A = L_0$ (and its complementary half is $B = R_0$):
\begin{equation}\begin{aligned}
    \rho_{A,\delta_F} \propto & \sum_{m,n=-xM }^{xM} c_m c_n \tr_{B} \left(U^m \ket{\psi_0}\bra{\psi_0} U^{\dagger n}\right) \\
    \propto & \sum_{m,n} c_m c_n \sum_{\bm{s},\bm{s}'} \tr_{B} \left( U^m\ket{l_{\bm{s}} r_{\bm{s}}} \bra{l_{\bm{s}'} r_{\bm{s}'}} U^{\dagger n} \right)
\end{aligned}\end{equation}
A first observation is that for $m \neq n$, the projections of $U^m\ket{l_{\bm{s}} r_{\bm{s}}}$ and $U^n\ket{l_{\bm{s}'} r_{\bm{s}'}}$ onto subsystem $B$ are orthogonal to each other, unless there are no initial excitations in region $\bm{R}_{|m|}$ of $\bm{s}$ and in $\bm{R}_{|n|}$ of $\bm{s}'$, i.e. $\bm{s} \cap \bm{R}_{|m|} = \bm{s}' \cap \bm{R}_{|n|} = \emptyset$. Among all $2^{N/2}$ possible configurations in subsystem $B$, however, there are only $2^{|m|}$ configurations with empty $\bm{R}_m$, whose contributions to the RDM are exponentially small in system size when $M \ll N$. It is thus safe to drop these terms and the RDM is almost ``diagonal'' in $m,n$. The rest ``diagonal'' terms can be grouped into two parts, $m \ge 0$ and $m < 0$:
\begin{equation}\begin{aligned}
    & \rho_{A,\delta_F} \\
    \sim & \sum_{m\ge 0} c_m^2 \sum_{ \bm{p} \subset \bm{M}_m }2^{N-2|m|}\ket{\Psi_{L,m}} \otimes \ket{l_{\bm{p} - m}} \bra{\Psi_{L,m}}\bra{l_{\bm{p}-m} } \\
    + & \sum_{m<0} c_m^2 \sum_{\bm{p} \subset \bm{M}_m } 2^{N - 2|m|}\ket{\Psi_{L,m}} \ket{r_{\bm{p}+|m|}} \bra{\Psi_{L,m}}\bra{r_{\bm{p}+|m|}}
\end{aligned}\end{equation}
where 
\begin{equation}
	\ket{\Psi_{L,m}} = \frac{1}{2^{N/2-|m|}} \sum_{\bm{q} \subset \bm{L}_m \backslash \emptyset} \ket{l_{\bm{q}-m} r_{\bm{q}+m}}
\end{equation}
collects all possible non-empty configurations in the left region of the system $\bm{L}_m$. $2^{N-2|m|}$ comes from the different configurations of initial excitation in $\bm{R}_m$ and the normalization factor of $\ket{\Psi_{L,m}}$. Here we similarly dropped the configurations with empty $\bm{L}_m$, to ensure that the states $\ket{\Psi_{L,m}} \ket{l_{\bm{p}-m}}$ (and $\ket{\Psi_{L,m}} \ket{r_{\bm{p}+|m|}}$) are orthogonal to each other for different configurations $\bm{p}$ in the middle region $\bm{L}_m$. Since there are in total $2^{2|m|}$ different initial configurations in region $\bm{M}_m$ for each $m$, the eigenvalues of $\rho_{A, \delta_F}$ are approximately $2^{2|m|}$ copies of $\sigma_m = \tilde{c}_m^2 2^{-2|m|}$ for $-x  M \le m \le  x M$, where 
\begin{equation}
	\tilde{c}_m^2 = c_m^2 / \sum_{m} c_m^2 \approx \begin{pmatrix} 2 M^2 \\ M^2 - m \end{pmatrix}^2 \big/ \begin{pmatrix} 4M^2 \\ 2M^2 \end{pmatrix}
\end{equation}
are the normalized filter coefficients. The scales of the normalized filter coefficients are illustrated in \cref{fig:filter_coef}. For $M \gg 1$, one may use the Stirling formula to obtain the approximation
\begin{equation}
	\sigma_m \approx \frac{2}{\sqrt{2\pi}M} e^{-\frac{2m^2}{M^2}}2^{-2|m|}.
\end{equation}
It is Gaussian function with width $M / 2 = 1 / 2\delta_F$ and centered at $m = 0$.

\paragraph{Scalings of entanglement entropies of filtered states.}
\label{sec:scaling_entropy_floquet}
Having the eigenvalues of the RDM of the filtered state, we are now able to compute the entanglement entropies. The von Neumann entanglement entropy is given by
\begin{equation}\begin{aligned}
    & S_1 (\rho_{A,\delta_F})\\
    = & - \sum_{m = - xM}^{xM} 2^{2|m|}\sigma_m \log \sigma_m\\
    \approx & \log \frac{\sqrt{\pi}M}{\sqrt{2}} + \frac{\sqrt{2}}{\sqrt{\pi}M} \sum_{m=-xM}^{xM} (2|m| + 2\log e \frac{m^2}{M^2}) e^{-\frac{2m^2}{M^2}} \\
    = & \log \frac{\sqrt{\pi}M}{\sqrt{2}} + \frac{2\sqrt{2}}{\sqrt{\pi}} \int_{-x}^{x} \dInt y\, ( M |y| + y^2\log e)e^{-2y^2}   \\
    =  & \log M + \frac{2M \gamma_x }{\sqrt{2\pi}} + \varO(1)   \\
    =  & \log \frac{1}{\delta_F} + \frac{\gamma_x}{\sqrt{2\pi}}\cdot \frac{2}{\delta_F}  + \varO(1).
\end{aligned}\end{equation}
In the derivation we take the limit $M \to \infty$ to turn the summation into integral with variable $y = m / M$. The coefficient
\begin{equation}
	\gamma_x = 2 \int_{-x}^{x} \dInt y\, |y|e^{-2y^2} < 1
\end{equation}
for any $x > 0$. In the cosine filter expansion of Gaussian function \cref{eq:floquet_filter}, we are using a truncated sum with index  $ - xM \le m \le xM$. For an untruncated version with $-2M^2 \le m \le 2M^2$, $x = 2M \to \infty$, and the coefficient becomes $\lim_{x \to \infty} \gamma_{x} = 1$.

Analogously, when we are taking an untruncated cosine filter expansion, the R\'enyi-$\alpha$ entropy for $\alpha \neq 1$ can be computed as
\begin{equation}\begin{aligned}
    & S_{\alpha} (\rho_{A,\delta_F})\\
    = & \frac{1}{1-\alpha}\log\left[ \sum_{m=-2M^2}^{2M^2} \sigma_m^\alpha\cdot 2^{2|m|} \right]\\
    \approx & \frac{1}{1-\alpha} \log \left[ \left(\frac{\sqrt{2}}{\sqrt{\pi}M}\right)^{\alpha} I(\alpha) \right],
    \label{eq:renyi_floquet_toy}
\end{aligned}\end{equation}
Where the integral
\begin{equation}\begin{aligned}
    I(\alpha)= & 2M \int_0^{\infty} \dInt y\, e^{-2\alpha y^2}2^{-2 M(\alpha-1)y}\\
    = & \frac{\sqrt{2}M}{\sqrt{\alpha}}  \int_0^{\infty} \dInt v \, e^{-v^2 - \frac{2 M(\alpha - 1)\ln 2}{\sqrt{2\alpha} }v  } && \hspace*{-0.6em} \left[v = \sqrt{2\alpha} y\right]\\
    = & \frac{2\ln 2}{\alpha - 1} u e^{ u^2 }  \int_0^{\infty} \dInt v\, e^{-  \left(v + u\right)^2 } && \hspace*{-2.6em} \left[u = \frac{(\alpha-1)M\ln 2}{\sqrt{2\alpha}}\right]\\
    = & \frac{\sqrt{\pi} }{(\alpha - 1)\ln 2} u  e^{ u^2 } \mathrm{erfc}(u),
\label{eq:integra_alpha}
\end{aligned}\end{equation}
where the complementary error function is defined as
\begin{equation}
	\mathrm{erfc}(u) = \frac{2}{\sqrt{\pi}} \int_{u}^{\infty} \dInt u\, e^{-u^2}.
\end{equation}
\begin{itemize}
	\item When $\alpha > 1$, $u > 0$. In the large $M$ limit $\mathrm{erfc}(u)$ has asymptotic expansion 
	\begin{equation}
		\mathrm{erfc}(u) = \frac{e^{-u^2}}{u\sqrt{\pi}} \sum_{n = 0}^{\infty} (-1)^n \frac{(2n-1)!!}{(2u^2)^n} \simeq \frac{e^{-u^2}}{u\sqrt{\pi}},
	\end{equation}
	and hence $I(\alpha) \simeq {1} / {(\alpha - 1)\ln 2 }$. Therefore
        \begin{equation}\begin{aligned}
            S_{\alpha > 1} (\rho_{A,\delta_F})  \approx & \frac{\alpha}{\alpha-1} \log \frac{\sqrt{\pi}M}{\sqrt{2}} + \varO(1)\\
            = & \frac{\alpha}{\alpha - 1} \log \frac{1}{\delta_F} + \varO(1).
        \end{aligned}\end{equation}
	\item When $0 < \alpha < 1$, $u < 0$, and $\mathrm{erfc}(u) \to  2$. Thus 
	\begin{equation}\begin{aligned}
            S_{\alpha< 1} (\rho_{A,\delta_F}) & \approx \frac{1}{1 - \alpha} \log \left[  \left(  \frac{\sqrt{2}}{\sqrt{\pi} M}\right)^{\alpha}  \frac{\sqrt{\pi} }{(\alpha - 1)\ln 2} u  e^{ u^2 } \right]\\
            & = \log M + \frac{u^2 \log e}{1 - \alpha} + \varO(1)\\
            & = \log \frac{1}{\delta_F} + \frac{(1-\alpha)\ln2}{ 2\alpha}\cdot\frac{1}{\delta_F^2} + \varO(1)
        \end{aligned}\end{equation}
\end{itemize}

There are separate scalings in R\'enyi entanglement entropies for different choices of $\alpha$, which are also observed by \cite{Morettini2024} with Hamiltonian filters. When $\alpha = 1$, the von Neumann entanglement entropy has a leading linear term in $1 / \delta_F$ and a subleading logarithmic term. 
When $\alpha > 1$, the R\'enyi entanglement entropy is only logarithmic in $1 / \delta_F$, indicating that the filtered states are not \emph{fully} thermalized for any $1 / \delta_F = \mathrm{poly} (N)$. We will further discuss how thermal the filtered state is in Section~\ref{sec:filter_thermal}.

For $0 < \alpha < 1$, the R\'enyi-$\alpha$ entanglement entropy turns out to be quadratic in $1 / \delta_F$. This quantity more closely relates to the simulability of MPS~\citep{Schuch2008}, which adds to the argument of numerical difficulties for classical simulations. This however holds only for exact Gaussian filters. If we take a truncated sum in the cosine filter expansion ($-xM < m < xM $), in \cref{eq:integra_alpha} $\mathrm{erfc}(u)$ should be replaced by $\mathrm{erfc}(u) - \mathrm{erfc}(u + \sqrt{2\alpha} x)$. It won't change the result when $\alpha > 1$, whereas if $\alpha < 1$, 
\begin{equation}\begin{aligned}
    & \mathrm{erfc}(u) - \mathrm{erfc}(u + \sqrt{2\alpha} x) \\
    = &\mathrm{erfc}(- u - \sqrt{2\alpha} x) - \mathrm{erfc}( - u)\\
    \simeq & - e^{-(-u - \sqrt{2\alpha} x)^2 } / u \sqrt{\pi}.
\end{aligned}\end{equation}
Inserting it into \cref{eq:renyi_floquet_toy}, we get 
\begin{equation}
	S_{\alpha < 1} (\rho_{A,\delta_F}^{\mathrm{trunc}}) \simeq x \cdot \frac{2}{\delta_F} - \frac{\alpha}{1 - \alpha} \log \frac{1}{\delta_F}.
\end{equation}
Thus for a truncated filter, the R\'enyi entanglement entropy is still linear in $1 / \delta_F$ when $0 < \alpha < 1$. This result is different from \cite{Morettini2024}, where they are stick to exact Gaussian filters. We can recover the quadratic term in the limit $x \to \varO(M) = \varO(1 / \delta_F)$.

\subsection{Global random unitary model}
\label{appendix:random_mat}
In this appendix we give another example, which verifies our ansatz in a more extreme setup. Let us think about the filtered state generated by a global Haar random unitary $U$:
\begin{equation}
    \ket{\psi_{\delta_F}} = \sum_m \tilde{c}_m U^m \ket{\psi}.
\end{equation}

\begin{figure}
    \centering
    \includegraphics[width=.4\textwidth]{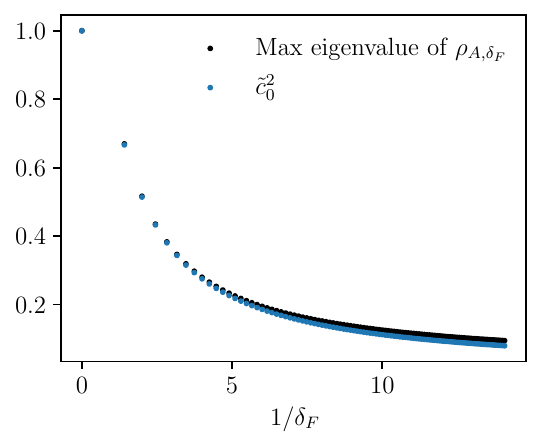}
    \caption{The largest eigenvalue of bipartite RDM of the filtered state generated by global random unitaries, compared with $\tilde{c}_0^2$.}
    \label{fig:largest_eigval}
\end{figure}

The arguments in \cref{sec:rdm_floquet} only need to be slightly modified. We can group the ``diagonal'' terms of the RDM $\rho_{A,\delta_F}$ into two parts: initial state $\tilde{c}_0^2 \rho_{A,00}$ and the rest thermalized terms $(1-\tilde{c}_0^2) \rho_{A, tt} = \sum_{m\neq 0}\tilde{c}_m^2 \rho_{A, mm}$, and the same steps follow. As a result, there will be an isolated large eigenvalue $\lambda_0\approx \tilde{c}_0^2$ (see \cref{fig:largest_eigval}), which corresponds to the contribution from 0-th order term.

\begin{figure}
    \centering
    \includegraphics[width=.235\textwidth]{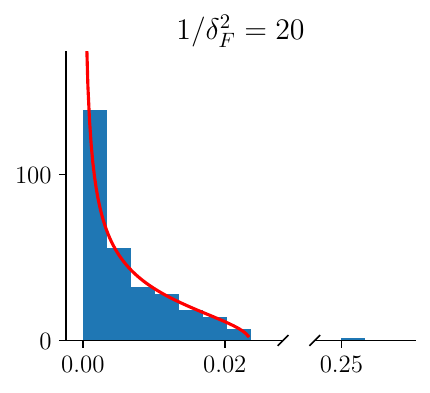}
    \includegraphics[width=.235\textwidth]{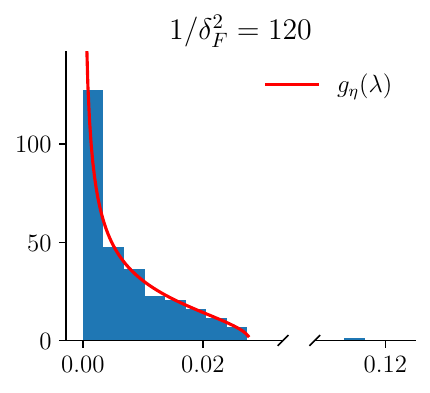}
    \caption{Density histograms of eigenvalues of the bipartite RDM of the filtered state generated by global random unitaries. The system size $N=14$, $N_A = 7$. Left: $1 / \delta_F^2 = 20$; right: $1 / \delta_F^2 = 120$. Red line: the density distribution $g_{\eta}\left(\lambda \right)$ as defined in \cref{eq:mp_dist}.}
    \label{fig:hist_eigval}
\end{figure}

For the rest of the eigenvalues $\left\{\lambda_i|0<i< 2^{N/2}\right\}$, our ansatz predicts that they should follow the (rescaled) Marchenko-Pastur distribution for the RDM of Haar random states, which has been thoroughly studied in random matrix theory~\citep{Nechita2007}. Let us define $s:= 2^{N_A} / 2^{N_B}$ to be the ratio of the subsystem Hilbert space dimension to the one of the environment, where we assume $N_B\ge N_A$ so that $0<s\le 1$. The probability density distribution $\left\{\lambda_{i}\right\}_{i>0}$ writes 
\begin{equation}
    g_\eta(x) = 
    \left\{
    \begin{aligned}
        &\frac{1}{2\pi s\eta} \sqrt{ \left(1- \frac{a\eta}{x}\right) \left(\frac{b\eta}{x} - 1\right)}, & a\eta \le x \le b\eta\\
        &0, & \text{otherwise;}
    \end{aligned}
    \right.
    \label{eq:mp_dist}
\end{equation}
where $a = (\sqrt{s}-1)^2 $, and $b = (\sqrt{s} + 1)^2$. The rescaling constant $\eta = (1 - \lambda_0) / (2^{N_A}-1)$ ensures that
\begin{equation}
    \begin{aligned}
        \tr (\rho_{A, \delta_F}) = \sum_{i=0}^{2^{N_A}-1} \lambda_i = \lambda_0 + (2^{N_A}-1)\int_{a \eta}^{b \eta} {g}_{\eta}(\lambda) \dInt\lambda =  1.
    \end{aligned}
\end{equation}
Specifically, if $s = 0$, then the RDM converges to the maximally mixed state.
Numerical simulations for small size systems support our arguments.
In \cref{fig:hist_eigval} the density histograms of RDM eigenvalues are plotted, which fit well with \cref{{eq:mp_dist}}.


R\'enyi-$\alpha$ entanglement entropies can now be computed:
\begin{equation}\begin{aligned}
    (1-\alpha)S_{\alpha}(\rho_A) = & \frac{1}{1 - \alpha}\log  \left[ \lambda_0^{\alpha} + \int_{a\eta}^{b\eta} \dInt \lambda\ \lambda^{\alpha} \cdot n\cdot g_{\eta}(\lambda) \right] \\
    = & \frac{1}{1 - \alpha} \log  \left[ \lambda_0^{\alpha} + (1 - \lambda_0)^{\alpha} {n^{1 - \alpha}}  I(\alpha, s) \right],
\label{eq:entropy_renyi}
\end{aligned}\end{equation}
where
\begin{equation}
    I(\alpha, s) = \frac{1}{2\pi s} \int_{a}^{b} \dInt u\, u^{\alpha}  \sqrt{\left( 1 - \frac{a}{u} \right) \left( \frac{b}{u} - 1\right)}    
\end{equation} 
is a constant independent of system size and $\lambda_0$. For example, $I(2,s) = 1 + s$ and $I(1, s) = 1$. 
    
When $\alpha > 1$, $n^{1 - \alpha}$ will be suppressed exponentially in system size, and thus for $\lambda_0 = \varO(\mathrm{poly}(1/N))$
\begin{equation}
    S_{\alpha}(\rho_A) \sim \frac{\alpha}{\alpha - 1} \log \lambda_0 \approx \frac{\alpha}{\alpha - 1} \log \tilde{c}_0^2. 
\end{equation}
For a finite size system, it will eventually saturate to volume law entanglement entropy $\sim N_A$, while before that $S_{\alpha>1}(\rho_A)$ first grows logarithmically with regard to $M$.

\begin{figure}
    \centering
    \includegraphics[width=.4\textwidth]{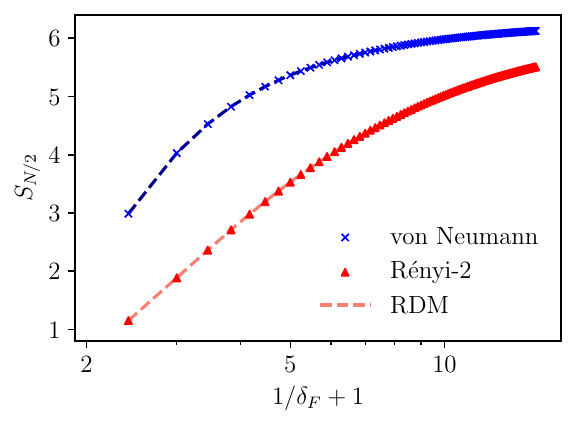}
    \caption{Von Neumann and R\'enyi-2 entanglement entropy of random filtered state. Here we only take one global random unitary as the sample, while its entanglement entropies are already close to the expectations due to typicality. The system size $N = 14$. We choose $s=1$, i.e., $N_A = N / 2$. RDM stand for \cref{eq:entropy_renyi} and \cref{eq:entropy_1} derived from eigenvalue distribution of random density matrices, taking $\lambda_0$ from the black scatter points of \cref{fig:largest_eigval}.}
    \label{fig:entropies}
\end{figure}

When $\alpha \to 1$, we can set $\alpha = 1 - \epsilon$ for a small value $\epsilon$ and obtain that
\begin{widetext}
    \begin{equation}\begin{aligned}
        S_{1-\epsilon}(\rho_A) & = \frac{1}{\epsilon} \log \left[ \lambda_0(1 - \epsilon \ln \lambda_0) +  (1 - \lambda_0)(1-\epsilon \ln(1 - \lambda_0))({1 + \epsilon \ln n})(1 - \epsilon \partial_{\alpha} I(\alpha,s)_{|\alpha=1}) + \varO(\epsilon^2) \right]\\
        & = ({1 - \lambda_0})\left[\log n  - \frac{ \partial_{\alpha} I(\alpha,s)_{|\alpha=1} }{\ln 2}\right] -\lambda_0 \log \lambda_0 - (1-\lambda_0) \log(1 - \lambda_0) + \varO(\epsilon).
    \label{eq:entropy_1}
    \end{aligned}\end{equation}
\end{widetext}
Therefore 
\begin{equation}
    S_1(\rho_A) = 
     (1 - \lambda_0) N_A + \varO(1).
\end{equation}
The dominant term is proportional to the subsystem size. For any fixed $M > 0$, we can already obtain volume law entanglement entropy. Both formulas are verified numerically with for system size $N=14$ ($n=127$), see \cref{fig:entropies}.

Finally, when $0 < \alpha < 1$, the dominating term would be $(1 - \lambda_0)^{\alpha} n^{1- \alpha} I(\alpha, s)$. Therefore
\begin{equation}
    S_{\alpha}(\rho_A) = N_A  +  \varO(1).
\end{equation}
It saturates to the maximal entanglement entropy, as the typical case of states directly evolved with global random unitaries.

\paragraph*{Alternative approach using Weingarten calculus.}
The R\'enyi-$\alpha$ entanglement entropy of the filtered state can also be exactly computed using Weingarten calculus for integers $\alpha \ge 2$. When $\alpha=2$, for example, the R\'enyi-2 entanglement entropy of a given state $\rho$ can be written as
\begin{equation}
    S_2(\rho_A) = - \log \tr \left[ X_A (\rho \otimes \rho) \right],
\end{equation}
where the operator $X_A$ swaps the $A$ partitions of the two copies of the state density matrix. For a filtered state, the doubled density matrix $\rho_{\delta_F} \otimes \rho_{\delta_F}$ has the form
\begin{equation}\begin{aligned}
        \rho_{\delta_F} \otimes \rho_{\delta_F}  = & \sum_{m,n,p,q}\tilde{c}_m\tilde{c}_n\tilde{c}_p\tilde{c}_q \\
            &U^{m}U^{p}\ket{\psi \otimes \psi}\bra{\psi \otimes \psi} (U^\dagger)^n(U^\dagger)^q,
    \label{eq:rho2}
\end{aligned}\end{equation}
where $\tilde{c}_m = c_m / \sqrt{\braket{\psi | P_{\delta_F}(U)^2 | \psi}}$ are the normalized coefficients.

To obtain the average of $S_2(\rho_A)$ over Haar-radom distributed $U$, we need to first  compute terms of the form 
\begin{equation}
    \braket{\psi \otimes \psi | (U^\dagger)^n(U^\dagger)^q X_A U^{m}U^{p}| \psi \otimes \psi}.
\end{equation}
For this purpose, one can use the Weingarten calculus~\citep{Roberts2017}. Only when the number of $U$ and $U^{\dagger}$ are equal to each other, the Haar random average can be nonzero. In this case, there exist different ways of pairing the left (respectively, right) leg of $U$ and the right (respectively, left) one of $U^{\dagger}$. For each choice of pairing, connect the corresponding legs and the unitaries $U$ and $U^{\dagger}$ can be removed. There will be an additional factor called the Weingarten function, depending on the permutation determined by the pairing. Finally we sum over all the different ways of pairing.

The calculation in our case can be significantly simplified. Note that with $k$ pairs of unitaries $U$ and $U^{\dagger}$, the corresponding Weingarten function has the scaling $O(1 / d^k)$, where $d=2^N$ is the Hilbert space dimension of $\rho$. For each pairing configuration, after cancellation of unitaries, two types of circuits can contribute factors other than 1: 
\begin{enumerate}
    \item  a bare loop that connects at least 4 legs of unitaries, that results in $d$;
    \item  self contractions of $X_A$ as listed below:
        \begin{equation}
            \centering
            \includegraphics{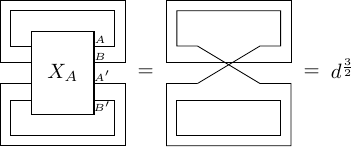}
        \end{equation}
        \begin{equation}
            \centering
            \includegraphics{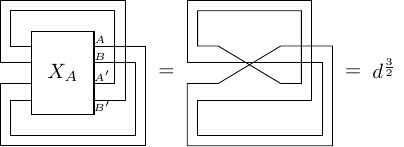}
        \end{equation}
        \begin{equation}
            \centering
            \includegraphics{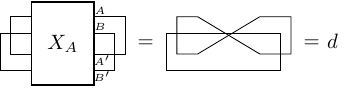}
        \end{equation}
        \begin{equation}
            \centering
            \includegraphics{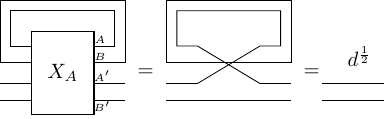}
        \end{equation}
\end{enumerate}
There are in total $4k$ legs of the unitaries, while $l$ of them connect to $X_A$ and another $l$ connect to $\ket{\psi}$, where $l$ is the number of nonzero integers in $m,n,p,q$. For the circuit to have nonzero average, $l \neq 1$. There can be at most $\lfloor (4k - 2l) / 4 \rfloor$ bare loops and $\lfloor l /2 \rfloor$ self contractions of $X_A$. Since each self contraction of $X_A$ contributes no more than $d^{3/4}$, in the thermodynamic limit the average of the trace for each term in \cref{eq:rho2} will vanish except in the case $m=n=p=q=0$, which gives $\tilde{c}_0^4$.

For the R\'enyi-2 entropy, we need to take the average only after taking the logarithm of the sum of the traces. Thus the Haar average should be taken after the Taylor expansion for $z = 1 - \tr \left[ X_A (\rho \otimes \rho) \right]$
\begin{equation}
    - \log (1 - z) = \sum_{s=1}^{\infty} z^s/s \ln 2, \quad 0 \le z < 1.
\end{equation}
With a similar argument as used above, only one term in $z$ contributes and finally we obtain
\begin{equation}
    S_2(\rho_{A, \delta_F}) = - \log \tilde{c}_0^4 \sim 2 \log \frac{1}{\delta_F}.
\end{equation}
The computation above indicates that in the large $N, M$ limit the entanglement entropy is only dependent on the zeroth order term coefficient $\tilde{c}_0$. It can be easily generalized to R\'enyi-$\alpha$ entropy with integers $\alpha>2$, which gives
\begin{equation}
    S_\alpha(\rho_{A, \delta_F}) \sim \frac{\alpha}{\alpha-1} \log \frac{1}{\delta_F}.
\end{equation}
This is the same as the results obtained from the eigenvalue distributions of their reduced density matrices.

\end{appendices}